\documentclass[12pt]{iopart} 
\usepackage{iopams}              
\usepackage[dvips]{graphicx}     
\usepackage{lscape}              
\usepackage{afterpage}           
\eqnobysec
\begin{document}
\jl{2} 
\topical{Geometry and symmetries of multi-particle systems}
\author[U Fano, D Green, J L Bohn and T A Heim]%
{U Fano\dag, D Green\ddag, J L Bohn\S\ and T A Heim$\|$} 
\address{\dag\ Physics Department \& James Franck Institute, 
University of Chicago, Chicago, IL 60637, USA}
\address{\ddag\ Department of Physics, P.O. Box 208120,
Yale University, New Haven, CT 06520, USA}
\address{\S\ JILA, University of Colorado, Boulder, CO 80309, USA}
\address{$\|$ Department of Physics and Astronomy, University of Basel,
CH--4056 Basel, Switzerland}
\begin{abstract}
The quantum dynamical evolution of atomic and molecular aggregates, from 
their compact to their fragmented states, is parametrized by a single 
collective radial parameter. Treating all the remaining particle coordinates 
in $d$ dimensions democratically, as a set of angles orthogonal to this 
collective radius or by equivalent variables, by-passes all 
independent-particle approximations. The invariance of the total kinetic 
energy under arbitrary $d$-dimensional transformations which preserve the 
radial parameter gives rise to novel quantum numbers and ladder operators 
interconnecting its eigenstates at each value of the radial parameter.

We develop the systematics and technology of this approach, introducing the 
relevant mathematics tutorially, by analogy to the familiar theory of angular
momentum in three dimensions. The angular basis functions so obtained are
treated in a manifestly coordinate-free manner, thus serving as a flexible
generalized basis for carrying out detailed studies of wave function evolution 
in multi-particle systems.  
\end{abstract}
\submitted 
\maketitle
\section{Introduction\label{Intro}}
The challenge for atomic and molecular theory to deal with progressively
larger aggregates of electrons and nuclei suggests treating them in terms of 
global parameters, in contrast to the usual independent-particle approach 
\cite{Macek1,Macek2,Fano81,Fano83a}. 
 Globally one may represent each configuration of $N$ particles 
by a single vector $\bi{R}$ identified by $3(N-1)$ internal coordinates with 
origin at its centre of mass.  $\bi{R}$'s modulus $R$ thus represents the
aggregate's overall size, while its direction $\hat{\bi{R}}$ specifies its 
geometry, i.e., the layout of the constituents' relative locations and 
orientations. Quantum dynamics controls then wave functions of $\bi{R}$ 
that emphasize $\hat{\bi{R}}$'s evolution from a compact to a fragmented
structure as $R$ grows.

In the centre-of-mass frame, the reduced two-body Coulomb problem separates
in spherical coordinates. Extending the system by adding more particles,
the \emph{kinetic} energy retains its three-dimensional spherical symmetry
as a subset of its symmetry in higher dimensions. This remark suggests
concentrating the search for appropriate coordinates on the kinetic energy.
Hyperspherical coordinates prove suitable for our task, by combining
the symmetries of particles' kinetic energies into a unified internal kinetic
energy of the whole aggregate. This procedure ensures an accurate description 
in the compact limit where the aggregate's kinetic energy predominates.

In a first step, hyperspherical coordinates separate $\bi{R}$'s modulus $R$ 
from its direction $\hat{\bi{R}}$, the latter being represented by parameters
analogous to the polar coordinates of physical space. The 
polar-coordinate symmetries, embodied in the familiar formalism of  angular
momentum theory, extend thus automatically to $\bi{R}$'s $(3N-3)$-dimensional
treatment.
Our proposed perspective of hyperspherical coordinates as extending
the three-dimensional spherical symmetry to higher dimensions affords 
advantages far beyond the technical aspect of providing a convenient coordinate
system. Separating the single hyper-radius $R$ from a large number of 
hyper-angles, and focusing on the symmetries under transformations of these 
hyper-angles, allows one to handle most of the many degrees of freedom in a 
multi-particle system analytically.  For example,
explicit calculation of angular integrals may be avoided in much the same 
way as in ordinary three-dimensional angular momentum theory by replacing such 
overlaps with coupling coefficients derived directly from the symmetry under 
($d$-dimensional) rotations.  A second example employs the \emph{%
coordinate-independent} representation of $d$-dimensional transformations
\cite{Cartan,Racah,Wybourne,Cornwell} 
 to construct complete sets of hyperspherical harmonics, higher-%
dimensional analogues of spherical harmonics, \emph{without} solving partial 
differential equations in any specific coordinate system. This application 
proceeds much as in three dimensions, where all harmonics may be built 
recursively from a single function once the appropriate ``laddering'' 
operators are identified.
The resulting flexibility in choosing centre-of-mass coordinate 
frames is crucial in the case of a multi-particle system with its evolving 
structure, because no single reference frame proves appropriate throughout the 
\emph{entire} evolution.

The hyper-radius $R$ serves as the ``evolution parameter'' of wave functions
$\Psi(R;\hat{\bi{R}})$, whose $\hat{\bi{R}}$-dependent features evolve with 
increasing $R$ toward their alternative fragmentation terminals. 
In more detail, the hyper-radius $R$ of an $N$-body aggregate of masses
$m_1, m_2, \ldots , m_N$, located at coordinates 
$\bi{r}_1, \bi{r}_2,\ldots, \bi{r}_N$ from the centre of mass is given by
\begin{equation} 
  R = \left( \sum_{i=1}^N \frac{M_ir_i^2}{\cal M}
\right)^{1/2}, \qquad {\cal M} = \sum_{i=1}^N M_i, \label{hyprad}
\end{equation}
In the special case of charged, point-like constituents
(electrons and nuclei) relevant to atomic and molecular applications.
the Schr\"odinger equation governing the aggregate's
evolution takes the hydrogen-like form,
in atomic units (a.u.) \cite{Macek1},
\begin{equation} 
\fl 
\left[ -\frac{1}{2{\cal M}}\left(\frac{\rmd^2}{\rmd R^2}+ 
\frac{\Delta_{\hat{\bi{R}}} }{R^2}\right) +
\frac{Z(\hat{\bi{R}})}{R} \right] R^{(3N-4)/2}
\Psi(R;\hat{\bi{R}}) = E\,R^{(3N-4)/2}\Psi(R;\hat{\bi{R}}).  \label{SEq}
\end{equation}
The factors $R^{(3N-4)/2}$, proportional to
the square root of a hyper-sphere's area with radius $R$, correspond to the
familiar factor $r$ of wave functions in physical space. Note how $(3N-4)/2$
reduces to unity for the H atom, where $N = 2$. Separation of these factors
affords non-zero values of $\Psi$ at $R=0$ and removes first derivatives from
\eref{SEq}. 

The general theory of hyperspherical coordinates (originally conceived for 
three-body scattering problems) dates to the fifties \cite{Morse,Fock,%
Wannier,Smith}. Since then, hyperspherical methods have been fruitfully 
applied to a wide variety of
many-body phenomena ranging from quantum chemistry to particle physics, as 
illustrated e.g.\ by several contributions in \cite{Tsipis}. 

Delves \cite{Delves1,Delves2} pioneered the method's application 
to shell-model calculations of nuclei. To this day, 
the hyperspherical approach remains a standard tool in nuclear physics, 
notably in the study of halo nuclei \cite{Cobis}, three-nucleon
systems \cite{Kievsky}, as well as large nuclei \cite{Gorbatov} and even 
sub-nuclear (quark) structure \cite{Santopinto}. (For reviews on 
hyperspherical methods in the context of nuclear physics, see 
e.g.~\cite{Smirnov,Fabre,Zhukov}.) 

In the seemingly very different 
context of reactive scattering in quantum chemistry, the participating 
atoms' coordinates have typically been cast in hyperspherical form, their 
motions being governed by effective potentials \cite{Kuppermann,Kaye1,Kaye2}.  
Besides molecular reactions 
\cite{Pack,Lindberg,Launay1,Launay2,Lepetit,Honvault,Groenenboom} the
hyperspherical approach applies also to molecular structure \cite{Yu} and
quantum phase effects in chemical reactions \cite{Wu,Adhikari}. 

These applications required more general and mathematical investigations on 
the structure and properties of hyperspherical functions 
\cite{Aquilanti1,Aquilanti2,Aquilanti3,Aquilanti4},
with a special focus on additional quantum-physical features such as
anti-symmetrization in fermion systems \cite{NKG88,Novoselsky,Barnea} 
and internal structures of an $N$-particle system in three-dimensional space 
\cite{Ovcharenko}. 

Applications of hyperspherical methods to doubly excited states of
two-electron systems in atomic physics \cite{Macek1,Macek2} resulted in 
a complete classification scheme for these states 
\cite{Lin81,Lin82a,Lin84a,Chen}, based on systematic investigations of
the wave functions' structure \cite{Lin82b,Lin82c,Lin83,Lin84b}. 
Connections with molecular structure \cite{Feagin,Watanabe86} arising 
from these classifications have more recently afforded extending the
hyperspherical approach to systems with several heavy particles 
besides electrons (i.e., to molecules) \cite{Tolstikhin}.  
In three-body Coulomb problems,
hyperspherical methods have become computationally competitive
through the ``hyperspherical close-coupling method'' \cite{Christensen},
and more recently 
in the form of ``diabatic-by-sector'' numerical methods \cite{Tang}.
Another extension employing over-complete basis sets afforded determining
accurately resonances of very high-lying doubly excited states close to the
threshold for double ionization \cite{HRS1,HRS2}.
Thus, the hyperspherical approach covers essentially the entire energy range
from the ground state, through the ``Wannier-region'' around threshold for
full disintegration, to energies high above this regime (see e.g.~\cite{Qiu}). 
Besides the two-electron atom and generic three-body Coulomb systems, 
studies have focused on doubly excited many-electron atoms  
\cite{Greene,Watanabe82}, with several forays into atoms with three 
\cite{Clark,Bao1,Bao2,Mori1,Mori2} or even more \emph{excited} electrons
\cite{Cavagnero1,Cavagnero2,Cavagnero3}. Moreover, the treatment of
highly excited atoms 
in external fields \cite{Schweizer1,Schweizer2} has extended the list
of successful applications of the hyperspherical method in yet another
direction.
Currently, physically adapted Sturmian basis sets promise further
advances in broader contexts \cite{AquilantiII}.
Reviews on various aspects of the hyperspherical approach in atomic physics
may be found, e.g., in \cite{Fano83b,Lin86,Lin95}. See also \cite{Chatto}
for a discussion of the reliability of the hyperspherical adiabatic method. 

Our equation 
\eref{SEq} implies going a step further, extending hyperspherical coordinates
to all constituents, electrons and nuclei, governed by their Coulomb
interactions. \Eref{SEq} thus represents an exact Schr\"odinger equation,
and all calculations proceeding from \eref{SEq} will be completely 
\emph{ab initio}
\cite{Fano81,Fano83a,Fano,Bohn1,Bohn2}.

We intend in this essay to encompass all types of applications of 
hyperspherical coordinates.  Thus the definition \eref{hyprad} need 
only extend over the dynamically relevant variables for a given
problem, ignoring, for instance, the coordinates of electrons belonging
to closed shells.  In this context note already the interplay between 
``motions'' on widely different (time) scales: 
At each fixed hyper-radius $R$, a ``geometrical'' structure emerges resulting 
from the faster motion in the hyper-angular coordinates;
the emerging structure then evolves on a different scale as $R$ increases.
This theme---central to the hyperspherical method---serves as a guideline 
throughout the present paper. As a qualitative illustration, consider the 
following hyperspherical description of a water molecule H$_2$O, a system 
consisting of ten electrons, two protons, and an oxygen nucleus. In a 
preliminary step towards \emph{ab initio} construction of this molecule, a
hyperspherical procedure would fill first the closed inner shells of the
constituent atomic cores, in this case only the $K$ shell of O$^{6+}$.
Recognizing the vast difference between electronic and nuclear motion,
the electronic motion is ``parametrized'' by the cores' arrangement. For 
each core arrangement, the electronic motion is analyzed to find the most 
favourable electron distribution. This analysis proceeds in a hyperspherical 
representation of the valence electrons as a single entity  consisting at 
the outset of a group of six electrons joined by two single electrons, as 
suggested by the core charges.
Analogous procedures for the construction and transformation of such 
``Jacobi trees'' will be outlined in \sref{CoordTrafo}. 
The particular arrangement of the 
cores should manifest itself in the hyper-radius $R_e$ associated with the set 
of valence electrons alone, once the latter are attached to the cores. The 
angular coordinates specifying the geometrical arrangement of the valence 
electrons depend parametrically on their own hyper-radius $R_e$, which is in 
turn parametrized by the ``size'' and ``shape'' of the nuclei (or cores).  The 
latter is itself described by a set of hyperspherical coordinates, 
$\{R_N, \alpha_N, \theta_N\}$, to be introduced in \sref{CoordTrafo}, thereby 
completing the 
hierarchy of geometrical structures governed by the interplay of 
hyper-radial and angular motions, and of electronic and atomic motions.  The 
dynamical evolution of the ``shape'' characterizing the cores' arrangement 
should minimize the energy of the whole molecule for the correct molecular 
geometry, while the hyper-angles characterizing the distribution of valence 
electrons should indicate that four of them are essentially attached to the 
O$^{6+}$ core, with the remaining two pairs forming the bonds holding the 
molecule together. Without attempting to carry out the computational details 
of this description, the present paper introduces techniques required for its 
implementation. Namely, we intend to focus on the universal 
(i.e. coordinate-independent) aspects of hyperspherical coordinates
and harmonics, to be implemented upon identifying the suitable Jacobi 
coordinates of a specific system.

Returning to equation \eref{SEq}, we note that features of $\Psi$ are 
discrete, owing 
to the finite extent of hyperspherical surfaces, being accordingly represented 
by appropriate quantum numbers and nodal structures rather than by coordinates,
much as they are in three dimensions; their interpretation will, however, 
require not only adequate mastery of high-dimensional geometry, a subject of 
this paper, but also of the dynamical interplay between light electrons and 
heavy nuclei. \Eref{SEq} should serve to calculate energy eigenfunctions 
for \emph{any} atom, molecule or analogous aggregate. Its solution for 
``collision complexes'' formed by colliding molecules will provide the 
relevant scattering matrix directly, as indicated in \sref{RadExp}, 
by-passing the 
calculation and study of energy surfaces.

The structure of \eref{SEq} parallels that of the atomic H equation, being
actually its extension to multi-particle systems. Its first term represents
the kinetic energy of the hyper-radial motions, its second term that of the
hyper-angular motions, and its last term the potential energy, which---being
the sum of Coulomb interactions among all of the aggregate's particle 
pairs---scales as $R^{-1}$. The evolution of $\Psi$'s  angular part as a 
function of $R$, foreign to hydrogen, stems from the non-zero value of the 
commutator $[\Delta_{\hat{\bi{R}}},Z(\hat{\bi{R}})]$.

\Eref{SEq} preserves instead hydrogen's invariance under coordinate 
rotations by securing invariance of the kinetic energy operator 
$\Delta_{\hat{\bi{R}}}/2 {\cal M}$ under rotations of $\hat{\bi{R}}$ about 
the centre of mass, by mass-weighting the
coordinates as described in the following. Sets of mutually independent 
eigenfunctions of $\Delta_{\hat{\bi{R}}}$'s, analogues of spherical 
harmonics called ``hyperspherical harmonics'', serve to expand \eref{SEq} 
into a 
system of coupled ordinary differential equations in the variable $R$, 
analogous to those of atomic physics. The ``effective atomic number'' operator 
$Z(\hat{\bi{R}})$ turns then into a matrix with rows and columns labelled by 
hyperspherical quantum numbers. 

Simple examples of such harmonics in $d$ dimensions have been formulated 
\cite{Avery,Erdely}, the corresponding eigenvalues of 
$\Delta_{\hat{\bi{R}}}$ having 
long been known. These examples, to be discussed in \sref{HypHarm}, provide 
solutions for the second-order 
eigenvalue problem of $\Delta_{\hat{\bi{R}}}$  in specific 
coordinate systems for $d$ dimensions.  Developing an appropriate systematics 
of multi-variable hyperspherical harmonics suited to each system (and thus 
coordinate-independent) constitutes, however, a major objective, to be 
approached in \sref{Classify} on the basis of symmetries alone.

The importance of such a description cannot be overemphasized. The distribution 
of $\bi{R}$'s multi-dimensional direction $\hat{\bi{R}}$ should 
represent specific features of each system flexibly, near and far from its 
centre of mass as well as at intermediate ranges. At short ranges, where the 
system is compact, this distribution should minimize the centrifugal effect 
represented by the eigenvalue of $\Delta_{\hat{\bi{R}}}$ in \eref{SEq}. 
At large ranges, where the system fragments, $\hat{\bi{R}}$'s distribution 
should represent 
alternative, mutually orthogonal, fragmentation channels. Both of these 
contrasting representations can be achieved in terms of flexible ``Jacobi 
(or centre-of-mass) coordinate'' sets, interrelated by algebraic 
transformations, the subject of \sref{JacobiSets}.

The present paper develops implications of the quantum discovery that pairs
of variables conjugate in Hamiltonian dynamics are actually related by Fourier 
transformations, whereby dynamics reduces to kinematics, more generally to 
\emph{geometry} (whence stems this paper's title). Expanding 
equation \eref{SEq} in hyperspherical harmonics reflects but one aspect 
of these implications; Jacobi coordinates provide a second aspect.

The angular Laplacian's eigenvalue in $d$ dimensions reads
$\lambda(\lambda+d-2)$, with integer $\lambda$. Thus,  
the system of coupled equations resulting from expansion of \eref{SEq}
into hyperspherical harmonics parametrized (in part) by the hyper-angular 
momentum $\lambda$ is formally infinite, owing to the infinite 
range of $\lambda$, and seems accordingly impractical. Key circumstances 
reduce, however, its size generally to a modest level:
\begin{itemize}
\item[a)]  For low-lying channels, the centrifugal term of 
\eref{SEq}, whose numerator rises as
$\lambda^2$, quenches the amplitude of $\Psi$'s components with large values
of $\lambda$ to negligible levels at small hyper-radius $R$. 
\item[b)] Correspondingly, its Coulomb term (prevailing at large $R$) has
eigenvalues similarly spread over many orders of magnitude, its lowest
one approximating the lowest dissociation threshold for molecules or ionization
threshold for single atoms and its highest one approaching the threshold for
full disintegration of the system.
\item[c)] The range of $\lambda$ values of practical relevance depends thus
critically on the energy range relevant to each step of evolution. So
do accordingly the dimensions of the corresponding set of hyperspherical 
harmonics and of the relevant $Z(\hat{\bi{R}})$ matrix. 
\end{itemize}

Early calculations \cite{Fano,Bohn1,Bohn2} have accordingly shown the range of 
$\lambda$ values relevant at each value of $R$ to be modest, thus 
affording ready numerical integration of \eref{SEq}. 
Each infinitesimal ``$\rmd R$'' step of this 
integration generates an infinitesimal rotation of the $\Psi(R;\hat{\bi{R}})$ 
wave function by $[Z(\hat{\bi{R}})/R] \rmd R$ in the $\hat{\bi{R}}$ space. 
The broad range of applications envisaged in the present article rests
on the power of its underlying recursive procedures and on the characteristic
aptitude of computer technologies to apply such procedures step by step.

The vector $\bi{R}$, representing the structure of a multi-particle 
aggregate and constructed by \emph{recursive} procedures, has been resolved 
above into its magnitude $R$ and direction $\hat{\bi{R}}$, corresponding to the 
aggregate's size and shape, respectively. The aggregate's shape, in turn, 
needs \emph{articulating} into appropriate parameters---multipole moments, for 
instance---representing structural features of each system, a task presenting 
a challenge to be approached in \sref{RadExp} in terms of equation \eref{SEq}'s 
eigenchannel solutions $\Phi_{\rho(R)}(\bi{R})$.
An elementary example of such developments is afforded by noting that an
aggregate's fragmentation elongates its shape, thus minimizing its moment of 
inertia about a symmetry axis. Analogous features should be systematically 
identified and utilized in specific  applications.

The following sections should introduce the reader to analytical tools 
serving to treat atoms and molecules of increasing size: (i) Coordinate systems 
whose dimensionality extends recursively and flexibly, adaptable to particle 
sets with different masses arising in molecular structure and collisions 
(\sref{CoordTrafo}); 
(ii) Prototype examples of hyperspherical harmonics suitable for multi-%
electron atomic systems with a single heavy centre (\sref{HypHarm}); 
(iii) Systematics 
of harmonics labelled by eigenvalues of commuting operator sets $\{H_i\}$, 
adapted to the evolving structures of atoms and molecules at increasing $R$ 
(\sref{Classify}). The final \sref{RadExp} will outline an analytical procedure 
to integrate \eref{SEq}, 
displaying the evolution of relevant wave functions and thereby 
casting the results of previous algebraic developments \cite{Wybourne,Cornwell}
 into a more
explicit geometrical framework. 

The wide range of symmetry applications, resulting from invariance under 
coordinate rotations and thus relevant for our application to quantum 
mechanics, is currently covered appropriately by \cite{Rau}.
Terminology drawn from \cite{Wybourne,Cornwell} will appear in ``\ldots'', but 
familiarity with these references is not assumed. 
\section{Linear coordinate transformations in higher dimensions
\label{CoordTrafo}}
Physical-space expansions into spherical harmonics hinge on their
geometrical and kinematical behaviour under coordinate rotations, labelled as
the ``angular-momentum theory''. Their extension to larger coordinate sets
affords flexibility for displaying geometrical and kinematical
features of multi-particle 
systems, at the price of extending and elaborating each system's treatment.
 
Whereas, in a three-dimensional prototype, rotations about the $x$-axis, i.e. 
in the $yz$-plane, are not independent of rotations about the $z$-axis 
because of involving the $z$-axis itself, in a multi-dimensional setting 
rotations about different axes are independent \emph{insofar} as they operate 
in \emph{separate} planes. Each elementary coordinate rotation is then 
properly identified, in multi-dimensional settings, as occurring within (or 
parallel to) the plane through a given \emph{pair of coordinate axes}, rather 
than as preserving a single invariant axis. Independent rotations thus occur in
\emph{non-crossing planes}, rather than within a single one, e.g., in the $xy$- 
and $zt$-planes in four dimensions. In $d$ dimensions the number of independent 
rotations is readily seen to equal the largest integer not exceeding $d/2$, 
i.e., $d/2$ for even values of $d$ and $(d-1)/2$ for odd $d$. This number, 
called the \emph{rank} of each transformation group \cite{Wybourne,Cornwell}, 
is usually indicated as $\ell = [d/2]$.

Independent Hermitian infinitesimal-rotation operators, corresponding to 
\begin{equation}  
l_z = -\rmi\left(x\frac{\partial}{ \partial y} - y\frac{\partial }{\partial x}
\right) = -\rmi\frac{\partial }{\partial \varphi}  \label{lz}
\end{equation}
in three dimensions, are indicated generically by $H_i$, here and 
in the following. Their analytic (differential or algebraic) expression, 
often analogous to \eref{lz}, depends on a group's structure and on
coordinate choices to be described later. The elementary example of the
helium atom, consisting of three particles and described in terms of six
coordinates $\{x_1,y_1,z_1,x_2,y_2,z_2\}$ with origin at its centre of mass,
involves three independent rotations represented, e.g., by
\begin{equation} 
\fl 
-\rmi\left(x_1\frac{\partial}{\partial y_1} - 
 y_1\frac{\partial}{\partial x_1}\right),
\qquad 
-\rmi\left(x_2\frac{\partial}{\partial y_2} - 
 y_2\frac{\partial}{\partial x_2}\right),
\qquad 
-\rmi\left(z_1\frac{\partial}{\partial z_2} - 
 z_2\frac{\partial}{\partial z_1}\right). 
\label{CartGen}
\end{equation}
The first two of these expressions are plainly analogues of single-particle's 
$l_z$; the interpretation of the last one---intermixing two particles'
coordinates---remains obscure at this point. Note also that finite rotations
are familiarly represented by exponential functions of infinitesimal
operators, as in the example of rotation by an angle $\varphi$ about a 
$z$-axis, represented by
\begin{equation}    
  \rme^{\rmi\varphi l_z}.  
  \label{lzef}
\end{equation}

Maximal sets of commuting operators $\{H_i\}$, such as the three operators in
\eref{CartGen}, perform in multi-dimensional settings the function performed 
by $l_z$ for three-dimensional rotations; their eigenvalues are also 
integers of either sign, often called \emph{``weights''}. After separating
the centre-of-mass motion of an $N$-particle aggregate, its $3N-3$ internal
coordinates are partitioned typically in 
three-dimensional sub-sets of \emph{single-particle} coordinates. Sub-sets of
independent infinitesimal rotation operators are then 
conjugate to angles $\varphi_i = \tan^{-1}(y_i/x_i)$
ranging from $0$ to $2\pi$, and thus effectively boundless. 
These three-dimensional coordinate sub-sets may also include angles 
$\theta_i = \tan^{-1}((x_i^2+y_i^2)^{1/2}\,/z_i)$, 
$\theta_i$'s range being restricted by centrifugal potentials near each of its
poles. Angular coordinates representing $\tan^{-1}(r_i/ r_j)$ of 
single-particle distances from a centre of mass are similarly confined by 
boundary conditions. For purpose of illustration, 
replace the Cartesian coordinates of 
helium with six (frequently used) hyperspherical coordinates, namely,
\begin{equation} 
\eqalign{
R = (x_1^2 + y_1^2 + z_1^2 + x_2^2 + y_2^2 + z_2^2 )^{1/2}, 
\\ 
\alpha=\tan^{-1}\left(\frac{x_2^2 + y_2^2 + z_2^2}{x_1^2 + y_1^2+
  z_1^2 }\right)^{1/2}, \\
  \theta _i = \tan^{-1} \left(\frac{(x_i^2 + y_i^2)^{1/2}}{z_i}\right), 
\\ 
\varphi_i = \tan^{-1} \left(\frac{y_i}{x_i} \right), 
}
\qquad i = 1,2. 
\label{HScoord}
\end{equation}
These definitions map the indistinguishability of electrons onto a
\emph{novel symmetry} under the reflection $\alpha\rightarrow\pi/2-\alpha$. 
The operators \eref{CartGen} take now the form, analogous to the last
expression in \eref{lz},
\begin{equation} 
  -\rmi\frac{\partial}{\partial \varphi_1},\qquad -\rmi\frac{\partial 
}{\partial \varphi_2},\qquad  -\rmi\frac{\partial}{\partial \tan^{-1} 
(\cos \theta_2 \sin \alpha / \cos \theta_1 \cos \alpha )}. 
\label{HSGen}
\end{equation}

Note that sets of commuting operators $\{H_i\}$ are subject to coordinate
transformations among equivalent sets. They are, in fact, suited to represent
invariants of relevant particle sub-systems. Each of these sets in $d$ 
dimensions is complemented by a much larger set of (generally) non-commuting 
operators, analogues of $l_x \pm\rmi l_y$ in three dimensions that raise or 
lower the $m$ quantum number of spherical harmonics $Y_{lm}(\theta,\varphi)$, 
respectively. Thereby one reaches the total of $d(d-1)/2$ (increasing
quadratically with $d$) unrestricted linear infinitesimal transformations in 
$d$ dimensions. Most of these operators characteristically involve coordinates 
of different particles, thus correlating their motions. E.g., combinations of 
equation \eref{lz}'s analogues 
\begin{equation}  
\fl 
J_{ij}^{xy} = 
-\rmi\left(x_i\frac{\partial}{\partial y_j} - 
 y_j\frac{\partial}{\partial x_i}\right),
\qquad 
J^{xx}_{ij} = 
-\rmi\left(x_i\frac{\partial}{\partial x_j} - 
 x_j \frac{\partial}{\partial x_i}\right),
\qquad i \neq j, 
\label{SONGen}
\end{equation}
would raise and/or lower eigenvalues of operator pairs $\{H_i,H_j\}$. [The
upper indices of the $J$ symbol denote the physical-space components, while its 
lower labels signify particles' indices, together specifying the relevant
variable pair in the $d$-dimensional space.] 

This and analogous considerations enlarge the scope of our study considerably,
yielding a total of $d(d-1)/2 = (3N-3)(3N-4)/2$ infinitesimal operators for 
$N$ particles in $d=3(N-1)$ dimensions. The resulting wealth of 
operators will be introduced here and developed
later.
As each $H_i$ involves two coordinates, odd values of $d$ imply that one
coordinate, often labelled by 0, fails to be included in any of the $\ell$
$H_i$'s, even though contributing to the set \eref{SONGen}.

To establish contact with the relevant mathematical theory of Lie groups
and Lie algebras \cite{Wybourne,Cornwell,Rau}, we observe that 
the set of linear infinitesimal rotation operators in $d$ 
dimensions considered in this section forms the ``special orthogonal group'' 
of transformations, designated as $SO(d)$; the term ``special'' referring to 
exclusion of dilations. Quantum mechanics extends this group to include the 
analogous complex transformations forming the unitary group $SU(d)$ of
transformations that preserve the complex Hermitian form $\sum_k x_k x_k^*$
instead of its real quadratic analogue. The full unitary group $U(d)$
results from adding multiplication with a complex phase $\exp(\rmi\alpha)$ to 
its ``special'' version $SU(d)$. The real part of the Hermitian form is 
preserved by orthogonal transformations, its imaginary part by ``symplectic'' 
transformations that preserve bilinear forms \emph{odd} under permutations of 
their elements, such as the spin invariant 
$u_{1/2}u_{-1/2}^*-u_{1/2}^*u_{-1/2}$.
\subsection{Symmetry under rotation reversal; ladder operators
\label{LadderOps}}
Reversal of a rotation's direction is represented, e.g.\ in \eref{lz}, by
switching the sign of the coordinate $y$ or of the imaginary unit, thereby
reversing the sign of the operators $l_z$ and of their eigenvalues $m$.
Symmetry under this reversal has been high-lighted in
\cite{Wybourne,Cornwell} by
replacing pairs of coordinate labels $(x_i,y_i)$ with pairs $(x_i, x_{-i})$,
thus replacing $y_i$ in \eref{SONGen}. The index $i$ runs thus over
$(1,2\ldots,\ell$), being complemented by an $x_0$ for spaces of odd
dimension $d$. Spherical coordinates include
then $\ell$ azimuths $\varphi_i$, with odd parity under rotation reversal, 
and $\ell$ polar angles $\theta_i$, changed by this reversal into 
$\pi-\theta_i$. 

Here we preserve the familiar notation with coordinate pairs
$(x_i,y_i)$, pointing out that the rotation reversal is often complemented with
the reflection through the coordinate plane $xz$, which automatically 
reverses the sign of $y_i$. The resulting combination reverses the handedness
(``chirality'') of each particle's space coordinates $(x,y,z)$.

Note first that, whereas the physical space operators $(l_x,l_y)$ change
by unity the eigenvalues $m$ of $l_z$, each of the $N$-particle 
operators \eref{SONGen} shifts the eigenvalues $\{m_i,m_j\}$ of an operator
\emph{pair} $\{H_i,H_j\}$. Whereas combinations $(l_x \pm\rmi l_y)$ act as 
``ladder'' operators raising or lowering the eigenvalues $m$ of $l_z$ by unity,
combinations of \emph{four} among these $N$-particle analogues \eref{SONGen}
raise or lower
eigenvalue \emph{pairs} simultaneously. The resulting rather elaborate
classification of operators became a central feature of the algebraic 
treatment \cite{Wybourne,Cornwell}; we shall follow a more direct approach. 

Recall how the ladder operators of physical space, $l_x\pm\rmi l_y$, emerge
as non-Hermitian combinations of the Hermitian pair $(l_x,l_y)$. Similarly,
non-Hermitian ``raising'' and ``lowering'' conjugate operators, designated 
here generically as $(a^{\dag}, a)$, respectively, may be viewed as 
combinations of two pairs of Hermitian operators, $a^{\dag} + a$ and 
$\rmi(a^{\dag} - a)$, symmetric and antisymmetric, respectively. 
Recall also that the physical-space Hermitian operators $(l_x,l_y)$ are 
antisymmetric and symmetric, respectively, under reflection through the 
$xz$-plane, according to standard ``Condon-Shortley'' conventions. 
A heuristic approach to constructing ladder operators might thus start by 
identifying combinations of operators \eref{SONGen} that are symmetric and
antisymmetric, without resorting to Lie-algebra procedures. 

Consider now the example of infinitesimal operators pertaining to the
four-coordinate set $\{x_i,y_i,x_j,y_j\}$. Besides the commuting pair
$\{H_i,H_j\}$, denoted here as $\{(x_i,y_i),(x_j,y_j)\}$, the set of operators  
includes the ones from \eref{SONGen}, labelled 
$(x_i,y_j)$  and $(x_i,x_j)$, respectively, as well as 
the additional two 
$(x_j,y_i)$, $(y_i,y_j)$, totaling six 
operators, four of which intermix $i$ and $j$ coordinates. This
last sub-set gives rise to two pairs of symmetric and antisymmetric Hermitian
operators, and thence to two pairs of desired non-Hermitian operators, one of
them ``raising-raising'',  designated as $(++)$, together with its conjugate
$(--)$, and ``raising-lowering'' operators 
$(+-)$ and $(-+)$. Commutators among these operators generate other 
operators of the set, much as commutators among the $\bi{l}$ components do, 
as detailed in \ref{ConstrLadder}. This particular example, with ladder
operators appearing simply as combinations of two components acting on 
$m_i$ and $m_j$ in the way familiar from $SO(3)$, is not generic for $SO(d)$.
Instead its structure is due to the well-known feature of 
orthogonal groups that $SO(4)$ factors into a pair of $SO(3)$ sub-groups. 
A  related example familiar in physics is afforded 
by the factoring of the (proper) Lorentz group $SO(3,1)$ 
into $SU(2) \times SU(2)$, with ``spinors'' of opposite chirality 
\cite[Chapter~11]{Rau}. 
Reference \cite{Wybourne} derives these results algebraically. 

Returning once again to the physical-space operators $l_x \pm\rmi l_y$,
recall how they transform elements of a spherical harmonics set $\{Y_{lm}\}$
into one another. They serve further to identify a set's range, e.g., by
causing its ``highest weight'' element $Y_{ll}(\theta,\varphi)$ to vanish when 
acted upon by $l_x+\rmi l_y$. The ladder operators outlined above will 
perform an 
analogous role for the much larger and multi-faceted sets of hyperspherical 
harmonics. To this end one may combine the operators \eref{SONGen} first 
into pairs symmetric and antisymmetric in their $(i,j)$ indices,
\numparts 
\begin{equation}  
J_{ij}^{xy\pm} = \frac{1}{\sqrt{2}}\left(J_{ij}^{xy}\pm J_{ji}^{xy}\right)
= \frac{1}{\sqrt{2}}\left(J^{xy}_{ij}\mp J^{yx}_{ij}\right),  
\label{ijsym}  
\end{equation}
since $J^{yx}_{ji} = - J^{xy}_{ij}$. These Hermitian operators are then paired 
into non-Hermitian operators, 
\begin{equation} 
J_{ij\pm}^{xy} = J_{ij}^{xy+} \pm\rmi J_{ij}^{xy-}.  
\label{ijpm}  
\end{equation}
\endnumparts 
The heuristically introduced operators of the type \eref{ijpm} fail to 
act correctly as ladder operators. Nevertheless,  
 examples of ladder operators with a similar structure arising as
superpositions of \emph{four} operators \eref{SONGen} will appear in 
(\ref{SO6GenCart}--$c$), playing a key role in \sref{Classify}. 
[Combinations \eref{SONGen} that include instead the 
unpaired variable $x_0$ of odd-dimensional systems do not lend themselves to 
the symmetrization \eref{ijsym}, being thus more nearly analogous to the 
three-dimensional $(l_x, l_y)$.] 

The numbers of infinitesimal operators \eref{SONGen} and of the resulting 
ladder
operators increase quadratically with the numbers of particles and of the
corresponding coordinates. It will turn out in following sections, however,
that a number $\ell$ of \emph{linearly independent} ``raising-lowering''
operator \emph{pairs}, equal to the number of commuting operators $H_i$, 
suffices
to generate complete orthogonal sets of hyperspherical harmonics. Each of
those sets corresponds to a choice of relevant coordinates and of the
$\{H_i\}$ set. 

The coordinates and their infinitesimal rotations, developed thus far in 
this section, would intermix in their dynamical 
applications with the
inertial effects of the mass-differences among various particles. These
complications can, however, be removed by appropriate mass-weighting of
the coordinates as anticipated in \sref{Intro} and implemented next.
\subsection{Jacobi coordinates\label{JacobiSets}}
Mass-weighting of coordinates has served in \sref{Intro} to define the
hyper-radius $R$ conveniently, contrasting it with the angular coordinates
represented by $\hat{\bi{R}}$. Analogous devices serve to weight appropriately 
the components of $\hat{\bi{R}}$ pertaining to particles with different masses 
$M_i$, making them homogeneous, and thus removing the Laplacian 
$\Delta_{\hat{\bi{R}}}$'s
explicit dependence on single-particle masses. To this end, generic sets of
``Jacobi coordinates'' have been introduced long ago \cite{Smith}, replacing 
the single-particle $\bi{r}_i$ by vectors $\bxi_i$ with the 
mass-weighted dimension \emph{mass$^{1/2}$-length}.

Alternative sets of Jacobi coordinates occur, reflecting alternative
groupings of particles, properly weighted by the mass of each group through
linear transformations with \emph{dimensionless} coefficients. These sets are
in turn interconnected by dimensionless linear transformations, each of whose
steps amounts to \emph{rotation in one plane}. Handling of Jacobi coordinates
becomes thus laborious even though each step be elementary.

Jacobi coordinates prove essential by identifying each fragmentation
channel through a particular \emph{Jacobi tree} separating at its base into
two branches corresponding to the relevant fragments. One thus displays the
evolution of each particle-aggregate toward a specific fragmentation channel
by the structure of relevant tree-shaped Jacobi coordinates. The label
``Jacobi tree'' reflects the evolution of a multi-particle system fragmenting
(i.e., ``branching out'') from a trunk into separate systems. [Developing a 
multi-particle wave function toward one among its alternative fragmentations 
presents instead an upside-down view of that tree.]

We show below a few simple prototype Jacobi trees, whose upper endings
correspond to single-particle labels. Permutation of two particles is 
represented by
rotating $(a)$ about its trunk. Changing of $(b)$ into $(c)$, often referred
to as \emph{transplanting} of branch 2, corresponds to the prototype 
transformation
of hyperspherical harmonics. Generic transformations resolve into sequences 
of permutations and transplantations. Tree $(e)$ is often called ``canonical''.
Multi-particle systems are represented by correspondingly articulated trees.

\setlength{\unitlength}{8.5pt}
\begin{center}
\begin{picture}(41,13)
\put(1,9){\line(1,-2){2.5}}
\put(0.8,10){1}
\put(5.8,10){2}
\put(6,9){\line(-1,-2){2.5}}
\put(3.5,2){\line(0,1){2.0}}
\put(2.8,0.3){$(a)$}
\multiput(8.5,1.25)(0,1){10}{\line(0,1){0.5}}
\put(11,9){\line(1,-2){2.5}}
\put(13.5,9){\line(-1,-2){1.25}}
\put(16,9){\line(-1,-2){2.5}}
\put(10.8,10){1}
\put(13.3,10){2}
\put(15.8,10){3}
\put(13.5,2){\line(0,1){2.0}}
\put(12.8,0.3){$(b)$}
\put(18,9){\line(1,-2){2.5}}
\put(20.5,9){\line(1,-2){1.25}}
\put(23,9){\line(-1,-2){2.5}}
\put(17.8,10){1}
\put(20.3,10){2}
\put(22.8,10){3}
\put(20.5,2){\line(0,1){2.0}}
\put(19.8,0.3){$(c)$}
\multiput(25.5,1.25)(0,1){10}{\line(0,1){0.5}}
\put(28,9){\line(1,-2){2.5}}
\put(29.6,9){\line(-1,-2){0.8}}
\put(31.4,9){\line(1,-2){0.8}}
\put(33,9){\line(-1,-2){2.5}}
\put(30.5,2){\line(0,1){2.0}}
\put(27.8,10){1}
\put(29.4,10){2}
\put(31.2,10){3}
\put(32.8,10){4}
\put(29.8,0.3){$(d)$}
\put(35,9){\line(1,-2){2.5}}
\put(36.25,9){\line(-1,-2){0.625}}
\put(37.5,9){\line(-1,-2){1.25}}
\put(38.75,9){\line(-1,-2){1.875}}
\put(40,9){\line(-1,-2){2.5}}
\put(37.5,2){\line(0,1){2.0}}
\put(34.8,10){1}
\put(36.0,10){2}
\put(37.3,10){3}
\put(38.5,10){4}
\put(39.8,10){5}
\put(36.8,0.3){$(e)$}
\end{picture}
\end{center}

The Jacobi trees just introduced serve to characterize multi-particle
systems by their \emph{hierarchy of composition}, i.e., by indicating the
order in which particles are joined to form sub-complexes of the entire
aggregate. The labels thus refer to particle indices.
A similar concept re-appears in a further context: 
Analogous trees illustrate \emph{coordinate systems} and their 
transformations, their branches labelling appropriate angles in 
$\hat{\bi{R}}$'s decomposition \cite{Smirnov,Cavagnero2,Cavagnero3}. 
Rotation of tree $(a)$ by an angle $0\leq \varphi \leq 2\pi$
represents then a simple rotation about an axis. Variation of a coordinate
$0\leq\theta\leq\pi$ maps onto the angle between two branches. 

Alternative sets of Jacobi coordinates $\hat{\bi{R}}$ correspond thus to
different tree structures. This circumstance adds further elaboration to our
procedure, yet serving to display whole aggregates' evolutions. These aspects
have not been apparent in the initial applications of the present approach,
dealing with very few particles, even though necessarily underlying the
treatment of any multi-particle system. Each transformation of hyperspherical
harmonics resolves accordingly into a transformation from one to another set of
Jacobi coordinates and a transformation of the corresponding harmonics.

Note the ``hierarchical'' aspect of Jacobi-tree construction, which adds
particles \emph{sequentially}, contrasting with the ``democratic'' view of the
multi-particle coordinates leading to the \emph{quadratic} increase of the 
number of \eref{SONGen} operators as a function of the particle number $N$.

Once again: Each step dealing with Jacobi coordinates is elementary, but
the number and combinations of different steps are large, a characteristic
generally encountered in computer operations requiring adequate strategy and
planning. Such operations may properly articulate into successive phases. We
anticipate, for example in dealing with molecules, to build first each atom's
inner shells independently, by Cavagnero's procedure 
\cite{Cavagnero1,Cavagnero2,Cavagnero3},
combining later the resulting atomic ions with the residual atomic electrons.

Similarly, the following sections present first alternative
combinations of three particles, lying in a plane with their centre of mass,
into alternative \emph{pairs} of mutually independent collective coordinates
$\bxi$, followed by transformations among these pairs. Combinations of
larger particle sets into further $\bxi$'s will be dealt with next,
utilizing analogous procedures recursively. Even more extensive procedures
will hinge on experience in treating large multi-particle aggregates.
\subsubsection{A three-particle prototype\label{ThreePart}}
The positions $\bi{r}_i,\, i= 1,2,3,$ of a three-particle set identify a
plane where their centre of mass also lies. These positions are represented by
these co-planar vectors, but their \emph{internal} kinematics involves only 
two independent
Jacobi coordinate vectors, $\bxi$, with $3(N-1) = 6$ degrees of freedom.

The standard procedure for constructing Jacobi coordinate vectors
considers first the positions of two among these particles, $1$ and $2$ here,
with masses $M_1$ and $M_2$ and positions $\bi{r}_1$ and $\bi{r}_2$. These
input data combine into a first Jacobi vector weighted by the square root of
the pair's ``reduced mass'' $M_{12} = M_1M_2/(M_1+M_2)$, namely,
\begin{equation}   
  \bxi_{12} = \sqrt{M_{12}} (\bi{r}_1-\bi{r}_2),
\label{mwJac1}
\end{equation}
whose centre of mass lies in the plane at
\begin{equation} 
  \bi{r}_{12} = \frac{M_1\bi{r}_1+M_2\bi{r}_2}{M_1+M_2}.
\label{cm1}
\end{equation}

The next step combines the first pair, with mass $M_1+M_2$ and 
centre of mass position $\bi{r}_{12}$, with the third particle lying at 
$\bi{r}_3$. This step is performed in accordance with \eref{mwJac1} 
yielding the second Jacobi vector,
\begin{equation}  
  \bxi_{12,3} = \sqrt{M_{12,3}} (\bi{r}_{12} -\bi{r}_3),
\label{mwJac2}   
\end{equation} 
with the reduced mass
\begin{equation}  
  M_{12,3} = \frac{(M_1+M_2)M_3}{M_1+M_2+M_3 }.
\label{redm2}   
\end{equation}
[The comma-separated sub-scripts indicate the particle sub-complexes to
be joined.]

An additional feature relates the Jacobi vector \eref{mwJac2} to its 
alternatives
corresponding, e.g., to the permutation of indices 1 and 3, i.e., 
$(1,2,3)\rightarrow (3,2,1)$, yielding the two-dimensional vector rotation
\begin{equation} 
\{\bxi_{32},\;\bxi_{32,1}\} = \{\bxi_{12} \cos\beta - 
\bxi_{12,3}\sin\beta,\; \bxi_{12} \sin\beta + \bxi_{12,3} \cos\beta \},  
\label{rotJac}  
\end{equation}
by the angle
\begin{equation}  
  \beta = \tan^{-1} \sqrt{\frac{M_2(M_1+M_2+M_3) }{M_1 M_3}}.
\label{rotJacAngle}    
\end{equation}
Analogous \emph{kinematic rotations} correspond to cyclic permutations of 
indices.
\subsubsection{Extension to multi-particle aggregates\label{MultiPart}}
The formulation of equation \eref{mwJac2}, with elements from \eref{mwJac1},
has clearly recursive character. It implies that any pair of Jacobi vectors, 
$\{\bxi_p,\bxi_q\}$, representing two sub-aggregates of particles 
centred at $\bi{r}_p$ and $\bi{r}_q$ with masses $M_p$ and $M_q$, 
respectively, combines effectively into a single vector
\begin{equation}  
  \bxi_{pq} = \sqrt{\frac{M_pM_q}{M_p+M_q}}(\bi{r}_p-
\bi{r}_q).   
\label{mwJac}   
\end{equation}

Similarly, restructuring of any ``Jacobi tree'' diagram, which represents a
specific sequence of particle combinations forming an aggregate, resolves
into sequences of vector-pair $\{\bxi_p, \bxi_q\}$ rotations within a
plane, analogous to that represented by \eref{rotJac}. 
Such restructurings have been
discussed amply in \cite{Smirnov} under the name of ``timber 
transformations'', the
word ``timber'' being suggested by association with ``Jacobi tree''. The simple
underlying principle, stated in that reference, lies in the feasibility to
resolve any rotation in multi-dimensional spaces into a sequence of plane
rotations, a feature familiar for the three-dimensional rotations of physical
space. \ref{TrafoJacobi} exemplifies this procedure.

Transformations between different Jacobi trees prove highly relevant to our 
subject of atomic and molecular few-particle systems for the following reason. 
The Coulomb coefficient 
$Z(\hat{\bi{R}})$ results familiarly from contributions proportional to the 
reciprocal distances between the $N(N-1)/2$ particle pairs. Flexible
sets of Jacobi coordinates afford treating each of these distances as a single 
coordinate, to be combined with others, thus avoiding the familiar need to 
expand each term of $Z(\hat{\bi{R}})$ into a multipole series. Thereby 
determining Coulomb interaction matrix elements 
reduces to calculating integrals over $1/r$ rather than over  
$1/|\bi{r}_i-\bi{r}_j|$. 
\section{Sample hyperspherical harmonics\label{HypHarm}}
The familiar spherical harmonics, $Y_{lm}(\theta,\varphi)$, serve as 
tensorial base sets for $(2l+1)$-dimensional transformations induced 
by rotations of the physical-space coordinates. In multi-dimensional contexts 
analogous base sets of hyperspherical harmonics serve the same purpose. The 
name ``harmonics'' identifies them as eigenfunctions of the angular 
Laplacian operator 
$\Delta_{\hat{\bi{R}}}$ in the $(d-1)$-dimensional space of $\hat{\bi{R}}$ with
eigenvalues $-\lambda (\lambda + d - 2)$.

To understand the term $d-2$ in this eigenvalue formula, note 
first that it reduces to unity for $d = 3$ yielding the familiar
eigenvalue $l(l+1)$ of the squared orbital angular momentum. The unit in this
expression corresponds to the \emph{single} angular coordinate $\theta$ that
accompanies the angle $\varphi$ in polar coordinates. Its contribution to the
eigenvalue, namely $l$, corresponds to the ``zero-point energy'', ``$hl$'', 
of a unit-mass particle oscillating along the $\theta$ coordinate in the
centrifugal field generated by its rotation along $\varphi$ with quantum
number $l$. The occurrence of $d - 1$ dimensions for the vector $\hat{\bi{R}}$
raises the number of its coordinates, besides $\varphi$, from unity to
$d - 2$, thus accounting for the eigenvalue term $d - 2$.

As the pair of angles $(\theta,\varphi)$ identifies a direction of
physical space, an equal number of indices $(l,m)$ identifies a harmonic
belonging to a $(2l+1)$-dimensional set, with the magnetic quantum number
$m$ labelled as a ``weight'' and $l$ in the role of ``highest weight''. 
Extending 
this parametrization to $d$-dimensional spaces requires us to describe sets of 
hyperspherical harmonics $Y_{\lambda\bmu}(\hat{\bi{R}})$, where $\lambda$ 
replaces the ``highest weight'' $l$, the vector $\bmu$ represents a set of 
$d-2$ complementary labels, and $\hat{\bi{R}}$ a corresponding set of $d-1$ angles.

This extension provides a main tool for the quantum mechanics of
multi-particle systems, as indicated in \sref{Intro} anticipating the 
relevance of the hyperspherical harmonics and of their treatment in 
\cite{Fano,Avery}. Recall
that the position vector $\bi{R}$ of an $N$-particle set (in its centre of
mass frame) has $d = 3(N-1)$ dimensions. The wave functions $\Psi(\bi{R})$
of such a system, envisaged in \sref{Intro}, are conveniently expanded in
hyperspherical harmonics in analogy to expansions in spherical harmonics 
\cite{Fano}.
Their Schr\"odinger equation reduces similarly to a system of coupled 
ordinary differential equations in the ``hyper-radius'' $R$.

The symbol for hyperspherical harmonics, $Y_{\lambda\bmu}(\hat{\bi{R}})$,
replaces the index $l$ of spherical harmonics by the index $\lambda$ 
corresponding to the eigenvalue $-\lambda (\lambda + d - 2)$ of the 
$(d-1)$-dimensional angular Laplacian $\Delta_{\hat{\bi{R}}}$. This index also 
stands for the degree of the homogeneous ``harmonic'' polynomial products 
$R^{\lambda}Y_{\lambda\bmu}(\hat{\bi{R}})$. The second index $\bmu$ 
replaces the 
index $m$ of spherical harmonics with a corresponding set of $d-2$ parameters 
(the dimension of $\hat{\bi{R}}$ less 1) that identify a specific harmonic of 
degree $\lambda$.

Expanding wave functions of a multi-dimensional $\bi{R}$  utilizes 
``complete orthogonal sets'' of hyperspherical harmonics. Completeness is
achieved by extending the range of $\lambda$ adequately. Only a finite set
of harmonics proves, however, relevant at any finite value of $R$, higher
values being effectively excluded by a generalized centrifugal potential at
small $R$, as
noted in \sref{Intro}. [This potential, involving the $\lambda$ 
parameter, includes contributions from derivatives of variables corresponding 
to the $\alpha$ coordinate in \eref{HScoord} and representing the quantum 
mechanical resistance of particles to compression by boundary conditions.] As 
$R$ increases more and more hyperspherical harmonics start contributing to
the relevant wave functions, requiring adequate frame transformations to 
reflect the appropriate fragmentation channels, as outlined in \sref{RadExp}.
Orthogonality 
requires identifying, for each $\lambda$ value, an adequate set of 
vectors $\bmu$ corresponding to the simple set of integer 
values $|m| \leq l$ of the spherical harmonics and to the relevant set of 
$\hat{\bi{R}}$ components. Higher dimensionality implies here more 
elaborate sets of $\bmu$ vectors.

Generating these sets in $d$ dimensions, labelled by $(d-2)$-dimensional
vectors $\bmu$, relies in essence on separating the Laplacian's variables.
In the three-dimensional prototype $Y_{lm}(\theta,\varphi)$, the $m$ label 
arises as an eigenvalue of the $l_z$ operator representing the number of 
nodes of the 
corresponding eigenfunction $\sin^{|m|}\theta\rme^{\rmi m\varphi}$, 
whereas the 
remaining $l-|m|$ nodes pertain to the $\theta$ variable. For hyperspherical 
harmonics, sets of \sref{CoordTrafo}'s $\ell$ commuting operators $H_i$ 
 may provide 
corresponding quantum numbers $m_i$ and eigenfunctions. The residual 
$\lambda-\sum_i |m_i|$ nodes would then pertain to the $d-\ell-1$ residual 
variables, analogues of $\theta$.
    
Whereas alternative orthogonal sets of spherical harmonics pertain to 
alternative orientations of the $\hat{\bi{z}}$ coordinate axis in 
three-dimensional space, 
analogous sets of hyperspherical harmonics pertain to alternative selections of
$\ell$ commuting operators $\{H_i\}$, not necessarily based on a coordinate set
as they were in \sref{CoordTrafo}. 
 Whereas spherical harmonics $Y_{lm}(\theta,\varphi)$ 
depend on the ``longitude'' $\varphi$ with $|m|$ ``meridian'' nodes and on the 
``co-latitude'' $\theta$ with $l-|m|$ ``parallel'' nodes, sub-dividing their 
plots into separate ``lobes'', hyperspherical harmonics depend---for 
particular coordinates---on ``longitudes'' $\varphi_i$, each with $|m_i|$ 
nodes, and on the nodal distribution in the remaining 
coordinates. Additional components of $\bmu$ pertain to alternative 
partitions of $\lambda$ that delimit the range of the parameters $|m_i|$ as 
well as of additional coordinates. 
Note how the number of $\bmu$ components increases \emph{linearly} with
the number $N$ of particles, contrasting again with the operators 
\eref{SONGen}, whose number increases \emph{quadratically} with $N$.
    
Consider now how the features of hyperspherical harmonics bear on equation 
\eref{SEq}'s expansion: The value of $\Delta_{\hat{\bi{R}}}$ in its 
 centrifugal term
depends \emph{only} on the parameter $\lambda$ of each harmonic, whereas the
coefficient $Z(\hat{\bi{R}})$ of its Coulomb term resolves for an $N$-particle 
set into $N(N-1)/2$ terms $\sim R/|\bi{r}_i-\bi{r}_j|$, cast as matrices in 
the $\{\lambda\bmu\}$ basis. The flexibility afforded by selecting the 
$\ell$ 
operators $H_i$ and the remaining $d-\ell-1$ coordinates should serve here to 
avoid, or at least minimize, resorting to multipole expansion of each 
term.

We anticipate that coordinate rotations in $d$ dimensions transform
generally any harmonic $Y_{\lambda\bmu}(\hat{\bi{R}})$ into a superposition
of the harmonics of its whole orthogonal set with coefficients
$D^{\lambda}_{\bmu',\bmu}$ analogous to those that serve
to transform spherical harmonics and form a ``representation'' of the rotation
group. This feature rests on the multi-dimensional rotations' aptitude to
resolve into sequences of two-dimensional rotations, thereby affording us to
express any $D^{\lambda}_{\bmu',\bmu}$ coefficient in terms of more
familiar three-dimensional Wigner's Euler-angle functions $d^{(l)}_{m'm}$.
The occurrence
of alternative sets $\{Y_{\lambda\bmu}\}$ with the same $\lambda$ value 
reflects the multiplicity of the infinitesimal operator basis. Systematic 
classifications of alternative sets $\{Y_{\lambda\bmu}\}$, appropriate 
to the number and structure of their coordinates, shall rest on the sub-group
chains of the relevant group, as outlined below. With this background we 
describe now two sample harmonics' sets.
\subsection{A generic structure\label{JacobiPol}}
Laplacian equations for hyperspherical harmonics in $d$ dimensions,
\begin{equation} 
  [\Delta_{\hat{\bi{R}}} + \lambda (\lambda +d-2)]
Y_{\lambda\bmu}(\hat{\bi{R}}) = 0,     
\label{HSEq}  
\end{equation}
lend themselves to solution by separation of variables, since
$\Delta_{\hat{\bi{R}}}$ amounts to a sum of terms 
$f(\hat{\bi{R}})(\partial/\partial x_i) g_i(\hat{\bi{R}})
(\partial/\partial  x_i)$
with metric coefficients $f(\hat{\bi{R}})$ and $g_i(\hat{\bi{R}})$.  Their
prototype example is afforded by the Schr\"odinger equation for the
He atom, with $N=3$ (1 nucleus and 2 electrons), $d=3(N-1)=6$, and with 
three polar coordinates \eref{HScoord}, yielding \cite{Morse}
\begin{eqnarray}
\fl 
\Delta_{\hat{\bi{R}}} =  
\frac{1}{\sin^2\alpha\cos^2\alpha}\frac{\partial}{\partial
\alpha}\sin^2\alpha\cos^2\alpha\frac{\partial}{\partial\alpha} 
+ \frac{1}{\cos^2\alpha}
\left[\frac{1}{\sin\theta_1}\frac{\partial}{\partial\theta_1}\sin \theta_1
\frac{\partial}{\partial \theta_1} 
+\frac{1}{\sin^2\theta_1}\frac{\partial^2}{\partial\varphi_1^2}\right] 
\nonumber \\ 
+\frac{1}{\sin^2\alpha}\left[\frac{1}{\sin\theta_2}\frac{\partial}{
\partial\theta_2}\sin\theta_2\frac{\partial}{\partial\theta_2}
+\frac{1}{\sin^2\theta_2}\frac{\partial^2}{\partial\varphi_2^2}\right]
\nonumber \\
\lo= 
\frac{1}{\sin2\alpha} \left( 
\frac{\partial^2}{\partial \alpha ^2} + 4 
-\frac{\bi{l}_1^2}{\cos^2\alpha} - \frac{\bi{l}_2^2}{\sin^2\alpha} 
\right) \sin 2\alpha,
\label{SO6angul}   
\end{eqnarray}
whose form on the last line arises from renormalizing the harmonics
$Y_{\lambda\bmu}$ by the volume element $\sin 2\alpha$.

A familiar approach \cite{Macek1,Macek2,Fano81,Fano83a} to solving \eref{HSEq} 
with the angular Laplacian \eref{SO6angul}
assumes first $Y_{\lambda\bmu}$ to depend 
on $(\varphi_1,\varphi_2)$ through a factor 
$\exp(\rmi m_1\varphi_1+\rmi m_2\varphi_2)$, whereby each 
$(\partial/\partial\varphi_i)^2$ element of \eref{SO6angul} amounts
to $-m_i^2$.  Thereafter each of the square brackets in \eref{SO6angul} 
reduces to $-l_i(l_i+1)$, provided $Y_{\lambda\bmu}$ depends on each 
$\theta_i$ through the associate Legendre function 
$\sin^{m_i}\theta_i\; P_{l_i}^{m_i}(\cos\theta_i)$, with the relevant 
``highest weight'' $l_i$ thus limiting the range of $|m_i|$. The residual 
operator on the right of \eref{SO6angul} has then eigenvalues 
$-\lambda(\lambda +4)$, with $\lambda$ partitioned as $l_1+l_2+2n$, and with 
the ``Jacobi polynomial'' eigenvector 
$P_n^{(l_2+1/2,l_1+1/2)}(\cos 2\alpha)$, 
according to (22.6.4) of \cite{Tables}. [The factor 2 multiplying $n$ and 
$\alpha$ in these expressions stems from the exponents and coefficients in 
the functions of $\alpha$ in \eref{SO6angul}]. The eigenvector of \eref{HSEq}
reads thus
\begin{equation} 
\fl
Y_{\lambda\bmu}(\hat{\bi{R}}) = 
\cos^{l_1}\alpha \sin^{l_2}\alpha 
P_n^{(l_2+\frac{1}{2},l_1+\frac{1}{2})}(\cos 2\alpha) 
Y_{l_1m_1}(\theta_1,\varphi_1) Y_{l_2m_2}(\theta_2,\varphi_2), 
\label{HSatt} 
\end{equation}
with the expected five-component $\bmu \equiv \{n,l_1,m_1,l_2,m_2\}$,
and with the spherical harmonic factors $Y_{l_im_i}(\theta_i,\varphi_i)$.

The set of harmonics \eref{HSatt}, with a given value of $\lambda$, consists of
a number of elements $w$, the dimensionality of the relevant space.
This number depends on the number of alternative partitions of $\lambda$ into
$(l_1,l_2,n)$ consistent with the relation $\lambda = l_1+l_2+2n$, and on the
alternative $2l_i+1$ values of each $m_i$. The prototype example of 
$\lambda= 2$ leads to the partitions: $(2,0,0),\;(0,2,0),\;(0,0,1),\;(1,1,0)$ 
and, in turn, to $5+5+1+9=20$ harmonics. 
Different partitions of $\lambda$ yield hyperspherical harmonics with 
alternative nodal patterns, reflecting alternative sharing of rotational
kinetic energy in different modes.
The algebraic determination of $w$ is discussed 
in \cite{Wybourne,Cornwell,Rau} and outlined in \sref{Classify}. The general 
expression \cite{Cornwell} for $w$ reduces in the present case ($d$ = 6) 
to $(\lambda +3)(\lambda +2)^2 (\lambda +1)/12$.

Notice how the eigenvalue parameters $(l_1,l_2,2n)$ play the role of
``weights'' in \eref{HSatt}, each of them amounting to the ``highest weight''
for a sub-group of the rotation group $SO(d=6)$, thus contributing to
the ``highest weight'' $\lambda = l_1+l_2+2n$ of the $SO(6)$ harmonics
\eref{HSatt}.  The symmetry under the sign reflection $(m\rightarrow -m)$ 
(cf.~\sref{LadderOps})
manifests itself in these harmonics not only through the symmetry of each
factor $Y_{l_im_i}$, but also through the parity of the polynomial $P_n$,
which is even or odd for even or odd values of the ``pseudo weight'' $n$,
combined with interchange of its upper indices. Note
also how $n$ substitutes for the weight $m_3$, an eigenvalue of the third 
operator \eref{HSGen}, which fails to commute with the operators 
$\bi{l}_i^2$ even 
though commuting with $(\partial/\partial\varphi_1)$ and 
$(\partial/\partial\varphi_2)$.

The harmonics \eref{HSatt} provide, of course, a basis for expanding 
correlated electron wave functions in helium, in the form 
$\Psi(R;\hat{\bi{R}})$ 
of \sref{Intro}, with the direction 
$\hat{\bi{R}}\equiv\{\alpha,\theta_1,\varphi_1,\theta_2,\varphi_2\}$  
\cite{Fano,Bohn1}.  The expansion coefficients $\Psi_{n,l_1,m_1,l_2,m_2}(R)$ 
represent then desired features of the relevant eigenfunction 
$\Psi(R;\hat{\bi{R}})$.
This representation does, however, emphasize single-electron aspects of the 
state through its parameters $(l_i,m_i)$ rather than through the global 
features anticipated in \sref{Intro}.

Note, on the other hand, that the procedure presented above to construct the 
harmonics \eref{HSatt} applies \emph{recursively}
to $(N>3)$-particle systems, as should the previous comments, with the  
following qualifications: Extending \eref{HSatt} to 
$N>3$ particles involves $N-1$ spherical harmonics 
$Y_{l_im_i}(\theta_i,\varphi_i)$, initially independent of one another, 
and $N-2$ Jacobi polynomials
in $\cos2\alpha_j$ with the full set of $3N-4$ variables. The index pairs of 
these polynomials, replacing the single-particle $(l_i +1/2)$ of 
\eref{HSatt}, reflect however the \emph{total} angular momenta, $\bi{L}$ or 
$\bi{J}$, of paired particle sub-sets as determined by relevant 
\emph{hierarchical} additions of single-particle momenta \cite{Cavagnero2}.
Note also how the structure of the harmonics \eref{HSatt}, and of their 
higher-dimensional analogues, exploits the commutability of the base 
operators $\{H_i\}$ and of their corresponding sub-group structure, with a 
notable exception: The angle $\alpha$ as defined in 
\eref{HScoord} does not coincide with the $\arctan$ of the third operator 
$H_i$ of \eref{HSGen}.  
\subsection{An alternative structure\label{GegenbPol}}
The particular sub-group structure of \eref{HSatt} stresses the 
single-particle features of each system, contrary to \sref{Intro}'s emphasis. 
We turn now accordingly 
to \cite{Avery}'s alternative construction of harmonics that
emphasizes different elements, dealing with $d = 3N-6$ unspecified coordinates 
besides a spherical harmonic $Y_{lm}(\theta,\varphi)$, without reference to 
single-particle positions. This approach thus utilizes a single eigenvector
$\rme^{\rmi m\varphi}$ of a \emph{single} operator $H_i$, contrasting with 
the full set occurring in \eref{HSatt} and its extensions to $N>3$.

As a preliminary to complementing a single spherical harmonic $Y_{lm}$ with
additional variables, reference \cite{Avery} views the Legendre polynomial 
$P_l(\cos \theta)$, invariant 
under rotations about the $z$-axis, as the particular Gegenbauer
polynomial $C_l^{\alpha}(\cos\theta)$ with $\alpha = 1/2$ (as defined in 
\cite{Tables}, Table 22.6) extending it from 3 to $d = 3(N-1)$ dimensions 
by setting its parameter $\alpha$ at $d/2-1$, and replacing $l$ with 
$\lambda$ ($d=6$ in our example). 
It also replaces $\cos\theta$ with the scalar product of unit vectors 
$\hat{\bi{R}}\cdot\hat{\bi{R}}'$, where $\hat{\bi{R}}'$ corresponds to the
reference axis $\hat{\bi{z}}$. The hyperspherical harmonic 
\begin{equation} 
  C_{\lambda}^{d/2-1}(\hat{\bi{R}} \cdot \hat{\bi{R}}')  
\label{Gegenb} 
\end{equation}
is thus invariant under $d$-dimensional rotations of $\hat{\bi{R}}$ about
a fixed axis $\hat{\bi{R}}'$ (i.e., rotations labelled by a single parameter 
$\varphi$), as 
well as under rigid rotations of the pair $\hat{\bi{R}} \cdot \hat{\bi{R}}'$.  
Note that Gegenbauer (as Legendre) polynomials consist of only even or of only 
odd powers for even or odd values of $\lambda$ or $l$, respectively.  

As the spherical harmonics $Y_{lm}$ are generated from $P_l(\cos \theta)$
by operators $l_x \pm\rmi l_y$, harmonics $Y_{\lambda\bmu}(\hat{\bi{R}})$
are generated from \eref{Gegenb} by infinitesimal operator analogues of 
$l_x \pm\rmi l_y$, each of them combining a derivative, which lowers by unity 
the degree of
their operand, with a compensating factor linear in a coordinate, thus 
replacing a nodal line of the operand with a different coordinate's nodal line.
[Typically, the operator $l_x +\rmi l_y$, as applied to $Y_{lm}$ with 
non-negative $m$, is equivalent to 
$\rme^{\rmi\varphi}\sin\theta (\partial/\partial\cos\theta)$: 
Its partial derivative reduces $Y_{lm}$'s
polynomial dependence on $\cos\theta$ by one degree, thus removing one
parallel-line node, its first factor adds instead a meridian-line node 
implied by the vanishing of its real or imaginary parts. To trace 
out this effect more explicitly, recast the operator in the form
\begin{displaymath} 
\fl
l_x + \rmi l_y = -\rmi\left(y\frac{\partial}{\partial z} -z\frac{\partial
}{\partial y}\right) + \left(z\frac{\partial}{\partial x} - 
 x\frac{\partial}{\partial z}\right) 
= -(x + \rmi y)\frac{\partial}{\partial z} + z\left(\frac{\partial}{\partial x}
 + \rmi\frac{\partial}{\partial y}\right). 
\end{displaymath} 
The first term on the right of this expression, when applied to the function
$Y_{lm}(\theta,\varphi)$ with a non-negative $m$ value, raises its $m$ index
by one unit through the combined action of its two factors: The first factor,
$x+\rmi y = r\sin\theta\rme^{\rmi\varphi}$ combines with the same factor within
$Y_{lm}$, thus raising by one unit the number of its meridian-line nodes as
well as the exponent of its $\sin^m \theta$ factor. Its second factor 
$(\partial/\partial z)$, equivalent here to 
$r^{-1}(\partial/\partial\cos\theta)$, reduces by one unit the 
degree of $Y_{lm}$'s 
polynomial dependence on $\cos\theta$, thus suppressing one of its 
parallel-line nodes. The second term on the right of the operator's 
expression cancels instead the function $Y_{lm}(\theta,\varphi)$ by setting 
it to zero.] 

A direct analogue of the successive action of $l_x + \rmi l_y$ operators on 
$P_l(\cos \theta)$ would apply to the Gegenbauer polynomial \eref{Gegenb} 
a sequence of
corresponding operators acting on components of $\hat{\bi{R}}$. A novel aspect 
emerges, however, at this point, since $\hat{\bi{R}}$ has several components 
orthogonal to $\hat{\bi{R}}'$ acted upon by alternative operators in 
 alternative 
sequences. Such alternative sequences will occur in \sref{Classify}, whereas 
reference \cite{Avery}
avoids them by treating all $\hat{\bi{R}}$ components uniformly, besides their
physical-space sub-set 
$\hat{\bi{R}}\equiv\{\sin\theta\cos\varphi,\sin\theta\sin\varphi,\cos\theta\}$.

To this end, reference \cite{Avery} follows the frequent practice of 
complementing an initial two-dimensional component 
($\cos\varphi$ or $\sin\varphi$) by multiplying
it with the factor $\sin\theta$ and adding a further $\hat{\bi{R}}$ component 
$\cos\theta$, whereby $\hat{\bi{R}}$ retains its unit magnitude. Iterating this 
extension $d-3$ times yields the canonical set of $\hat{\bi{R}}$ components
\begin{equation}
\begin{array}{l}  
\cos\theta_1  \\
\sin\theta_1\cos\theta_2  \\
\vdots  \\
\sin\theta_1\sin\theta_2\cdots\cos\theta_{d-2}  \\
\sin\theta_1\sin\theta_2\cdots\sin\theta_{d-2}\sin\varphi  \\
\sin\theta_1\sin\theta_2\cdots\sin\theta_{d-2}\cos\varphi.
\end{array}
\label{AveryCoord} 
\end{equation}
For purposes of orientation consider that, if all $\hat{\bi{R}}$ components had
comparable magnitudes, each value of $\cos\theta_i$ would be of order $1/d$,
and hence each $\theta_i$  close to $\pi/2$. After elimination of
$R\cos\theta_1$, the factor $R\sin\theta_1$ of all successive components
would represent their total magnitude. The successive factors $\sin\theta_i$ 
contribute then to reduce the effective residual components of $\hat{\bi{R}}$
progressively.

The set of integer components of the vector label 
$\bmu$, non-negative and of decreasing magnitude, is similarly 
indicated by 
\begin{equation} 
  \bmu \equiv \{ \mu_1,\mu_2,...,\mu_j,...\mu_{d-1} \},
\label{GgbIndex}  
\end{equation}
with $\mu_{d-1} \equiv |m|$, $m$ being the multiplier in the harmonic's phase 
$m\varphi$.  The difference between successive components \eref{GgbIndex}, 
$\mu_j-\mu_{j+1} \geq 0$, represents the number of nodes in the $j$-th 
harmonic's dependence on $\cos \theta_j$, including the $|\mu_{d-1}|$ nodes 
implied by the phase factor $\rme^{\rmi m\varphi}$. The set of harmonics 
compatible with these specifications, for the same prototype values 
$\lambda = 2$ and $d = 6$ as for the harmonics \eref{HSatt}, has likewise
20 elements.

The resulting hyperspherical harmonic, equation (3.69) of \cite{Avery}, 
consists thus
mainly of products of $d-2$ Gegenbauer polynomials,
\begin{equation} 
  Y_{\lambda\bmu} = N_{\lambda\bmu} \prod_{j=1}^{d-2}
(\sin \theta_j)^{\mu_{j+1}}C_{\mu_j-\mu_{j+1}}^{\alpha_j+\mu_{j+1}}
(\cos \theta_j) e^{im\varphi}.  
\label{HSAvery}  
\end{equation}
The combined degree of this harmonic, equalling its total number of
nodes, amounts to the $\sum_{j=1}^{d-2}(\mu_j-\mu_{j+1}) = \lambda$, including 
the $\mu_{d-1} \equiv |m|$ nodes attributable to the $\rme^{\rmi m\varphi}$ 
factor.  
The $\alpha$ index of each Gegenbauer polynomial consists of two terms, 
$\alpha_j$ and $\mu_{j+1}$, the first of which, $\alpha_j = (d-j-1)/2$,
corresponds to the actual dimensionality of all the \eref{HSAvery}
factors with indices 
larger than $j$.  The second term, $\mu_{j+1}$, represents the additional 
effective dimensionality attributable to the factors 
$(\sin\theta_j)^{\mu_{j+1}}$
[whose combination with $C_{\mu_j-\mu_{j+1}}^{\alpha_j+\mu_{j+1}}$ amounts to 
an analogue of the associate Legendre function $P_l^m(\cos \theta)$].  The 
remaining factor of \eref{HSAvery}, $N_{\lambda\bmu}$, represents the 
normalization factor contributing to orthonormalize each set of harmonics 
\eref{HSAvery} with equal $\lambda$ and alternative $\bmu$ indices.

Each factor $(\sin\theta_j)^{\mu_{j+1}}$, which decays rapidly as
$\theta_j$ approaches a pole (at $0$ or $\pi$) corresponds to the factor 
$(\sin \theta)^{|m|}$ of associate Legendre functions, which reflects the 
centrifugal potential generated by its factor $\rme^{\rmi m\varphi}$.  
Indeed the exponent $\mu_{j+1}$ equals the total number of nodes in 
\eref{HSAvery} factors with indices larger than $j$, which contribute a 
centrifugal potential to the equation 
governing $C_{\mu_j-\mu_{j+1}}^{\alpha_j+\mu_{j+1}}$; each difference 
$\mu_j-\mu_{j+1}$ corresponds to a separation parameter $l_i(l_i+1)$ in the 
construction of the harmonics \eref{HSatt}.

In conclusion, the present construction of hyperspherical harmonics has 
followed two approaches: (i) Solving the (angular) Laplace equation by 
separation of variables, leading to the harmonics \eref{HSatt} and extensible 
to higher dimensions; (ii) Constructing the harmonics \eref{HSAvery} by a 
sequence of Gegenbauer polynomial 
factors (also separating variables), complemented by factors 
$(\sin\theta_j)^{\mu_{j+1}}$, analogues of the $(\sin\theta)^m$ of spherical 
harmonics and similarly generated by applying infinitesimal operators 
$\sin\theta (\partial/\partial\cos\theta)$ to the invariant Gegenbauer 
harmonic \eref{Gegenb}.

These approaches differ in their coordinates as well as in their dynamical 
implications: No physical specification of the $\hat{\bi{R}}$ components 
$\theta_j$ has occurred in the approach (ii). Approach (i) has relied on the 
single-particle coordinates and on their interrelations introduced in 
\eref{SO6angul}, namely 
$\{0\le\alpha\le\pi/2,\; 0\le\theta_i\le\pi,\; 0\le\varphi_i\le 2\pi\}$.
[The limited range of $\alpha$ coordinates reflects their definition through 
ratios of non-negative variables.]  Approach (ii) has utilized, in 
\eref{AveryCoord}, ratios of $\hat{\bi{R}}$ 
components restricted to real values of either sign, in addition to a 
\emph{single} complex phase, without explicit reference to single-particle 
coordinates (considered elsewhere in \cite{Avery}). The occurrence of 
two (or more) 
complex-coordinate phases in the approach (i) utilizes the commuting operator
set $\{H_i\}$, a dynamical element foreign to the approach (ii).  Intermediate 
approaches, utilizing that set partially, appear readily accessible.
\section{Classification and construction of hyperspherical harmonics
\label{Classify}}
\Sref{HypHarm} introduced hyperspherical harmonics for two different sets of
coordinates, relying on single-particle features to a different extent.
Both sets of hyperspherical harmonics are characterized as ``harmonic 
polynomials'',
i.e., as eigenfunctions of the angular Laplacian, obtained directly by
separation of variables. Their construction thus ties each of these harmonics 
to a particular choice of coordinates. 

By contrast, sections \ref{Intro} and \ref{CoordTrafo} 
repeatedly stressed the need for flexibility
in the choice of coordinates to describe the evolution of an atomic or
molecular complex from its compact to its fragmented states. Having 
familiarized the reader with properties of hyperspherical harmonics in the 
preceding
section, we now introduce harmonics essentially frame-independent, thus
by-passing extensive frame transformations necessary for harmonics tied to
a particular set of coordinates. We describe here a procedure to generate
complete sets of harmonics \emph{independent} of the choice of coordinates.
Consequently, the same set of harmonics serves throughout the entire
evolution process, frame transformations reducing to the task of expressing 
the harmonics in whichever coordinate system appears appropriate at any 
given stage of the evolution, without change to the basic structure of the 
functions themselves. 

The key feature enabling a definition of harmonics without separation of
variables in the Laplacian's eigenfunction equation rests on identifying the
Laplacian's symmetry under rotations in $d$ dimensions. \Sref{CoordTrafo}
described these transformations in terms of \emph{first order} infinitesimal 
rotation operators. We will now construct functions based on 
$\{H_i\}$-operators 
and on the corresponding ladder operators only. Because these functions will
possess the underlying symmetry of the Laplacian, they are ``harmonics''
\emph{a fortiori}.

The present task amounts to extending the \emph{classification} of spherical 
harmonics by their label pair $(l,m)$ to multi-dimensional systems, in
accordance with procedures to construct such harmonics. To this end recall 
that: 
\begin{itemize}
\item[a)] The $l$ label of $Y_{lm}$ identifies both an eigenvalue of the 
\emph{second-order} angular Laplacian $\bi{l}^2$ and the range 
$0\leq |m|\leq l$ of its second label.
\item[b)] The label $m$ itself is an eigenvalue
of the \emph{first-order} operator $l_z$.
\item[c)] Alternative equivalent sets of 
harmonics correspond to alternative orientations of the $z$-axis.
\item[d)] Complete 
sets of spherical harmonics $Y_{lm}(\theta,\varphi)$ emerge by operating on the
invariant harmonic $P_l(\cos\theta)$ with the conjugate pair of ladder 
operators $l_x\pm\rmi l_y$, or alternatively operating with the lowering 
$l_x -\rmi l_y$ alone on the ``highest weight'' harmonic 
$Y_{ll}(\theta,\varphi)$.
\end{itemize}
Corresponding remarks on hyperspherical harmonics outline here this
section's development:
\begin{itemize}
\item[a)] We have seen how the multi-dimensional label $\lambda$ performs $l$'s
role in identifying eigenvalues of the relevant multi-dimensional angular 
Laplacian. Alternative partitions of $\lambda$'s value will similarly delimit
the ranges of $m$'s analogues.
\item[b)] Analogues of $m$ are the integer (or half-integer) eigenvalues 
  $m_i$ of the $\ell$ first-order commuting operators $\{H_i\}$ introduced in 
\sref{CoordTrafo}. These sets are viewed as 
components of a vector $\bi{m}$ in the space subtended by the 
operators $\{H_i\}$. They were noted, however, in \sref{JacobiPol} 
not to be fully compatible with 
the corresponding parameter set $\{l_i\}$ that delimits each $\{|m_i|\}$'s 
range, equation \eref{HSatt}'s label $n$ replacing the eigenvalue $m_3$ of a  
different equation. Harmonic eigenvectors of the $\{H_i\}$'s will be
nevertheless identified by convenient sets of $\bi{m}$ components, i.e., by
lattice points in the $\ell$-dimensional $\{H_i\}$ space.
\item[c)] Alternative equivalent sets of hyperspherical harmonics correspond 
to alternative orientations of a vector $\blambda$ in the relevant 
$\{H_i\}$ space and to alternative analyses of a system's dynamics. 
Transformations of coordinates and/or of the $\{H_i\}$ set 
yield equivalent sets of hyperspherical harmonics.
\item[d)] Hermitian-conjugate ladder operators, analogues of physical-space's 
$l_x\pm\rmi l_y$, will emerge as superpositions $E_{\balpha}$ of 
Hermitian \emph{pairs}---as anticipated in \sref{CoordTrafo}---with vector 
labels $\balpha$, 
each label with unit-magnitude components in the $\{H_i\}$ space, being thus  
represented by diagonal vectors in that space. Linearly independent 
$\ell$-dimensional \emph{sub-sets} of these operators, denoted by 
$\{E_{\bfeta_s}\}\equiv\{E_{-\bfeta_s}^{\dagger}\}$, $s=1,\ldots,\ell$
 suffice for the present task \cite{Wybourne,Cornwell}, as well as 
for encompassing all the far more numerous 
ladder operators by their own appropriate superpositions.
\end{itemize}
A sample set of harmonics emerging in this framework will be displayed at
the end of \sref{ExampleSO6}.

Quantum numbers $m_i$, eigenvalues of $H_i$ operator sub-sets, serve to 
classify ``(quasi)-invariants'' of multi-particle systems. Such is the $J_z$ 
component of the invariant angular momentum $\bi{J}$ of isolated systems. Two 
$m_i$ components pertain to a molecule rotating about a symmetry axis of lower 
inertia in its ``body-frame''. Three $m_i$'s characterize atomic Rydberg 
states, one of them pertaining to their inner core, one to the Rydberg electron
and one to their vector sum. A plethora of such $m_i$'s might pertain to a 
turbulent fluid. 

In the absence of quasi-invariants, other than the total $\bi{J}$'s, the
single label $m$ of \eref{HSAvery} may suffice, but a number $(\leq\ell)$ of
additional $m_i$'s, judiciously chosen with reference to the system's
structure, helps classifying harmonics $Y_{\lambda\bmu}(\hat{\bi{R}})$,
and the corresponding multi-particle wave functions.

Geometrical elements of classification also emerge from the nodal patterns 
of harmonics, as noted in \sref{HypHarm}. The Laplacian's separability into 
coordinates $x_i$, stressed at the outset of \sref{JacobiPol}, 
 affords real solutions
of the several resulting one-dimensional equations in $x_i$ (with appropriate
boundary or periodic conditions) to be characterized by $n_i$ nodes, yielding
altogether $\sum_i n_i$ nodes, a total basically equal to the eigenvalue
$\lambda$. Real hyperspherical harmonics with equal $\lambda$ differ
then by their $\lambda$'s partitions into the relevant $n_i$.

For the specific purpose of constructing sets of hyperspherical harmonics,
sets of $E_{\pm\bfeta_s}$ operators, analogues of the physical-space 
``ladder operators'' $l_x\pm\rmi l_y$, complement sets of $H_i$ conveniently, 
much as the $l_x\pm\rmi l_y$ complement $l_z$. To this end, the total angular 
momentum's $J_z\equiv H_1$ may be complemented by appropriate $H_i$ and 
thence by corresponding ladder operators 
$\{E_{\bfeta_s}\}\equiv\{E_{-\bfeta_s}^{\dagger}\}$. 
In principle, quantum numbers $m_s$ (of either sign) of the 
desired harmonics represent then the number of $E_{\pm\bfeta_s}$ having 
acted on the invariant harmonic $C_n^{\alpha}(\hat{\bi{R}}\cdot\hat{\bi{R}}')$, 
\eref{Gegenb}. In practice, the derivation of hyperspherical harmonics
for a given $\lambda$ proceeds more appropriately  by acting on the analogue 
of the spherical $Y_{ll}(\theta, \varphi)$ with sequences of lowering operators
$E_{-\bfeta_s}$,
 since the ``highest weight'' hyperspherical harmonic is uniquely defined.

The $m_s$ values raised or lowered by $E_{\pm\bfeta_s}$ operators in 
this procedure are delimited by the relevant ``highest weight'' eigenvalue 
$\lambda$ of $J_z\equiv H_1$, viewed as a vector $\blambda$ directed 
along a particular axis and expanded as 
\begin{equation} 
  \blambda=\sum_s \lambda_s\bfeta_s.
  \label{lambdaPart}
\end{equation}
The partition coefficients $\lambda_s$ are integers (or half-integers) insofar
as both $\blambda$ and the $\bfeta_s$ are vectors of the $\{H_i\}$ 
space with coefficients of integer (or half-integer) magnitude. The $\lambda_s$
values serve thus as ``highest weights'', analogues of \eref{HSatt}'s $l_i$, 
for the $m_s$ quantum numbers. The $m_s$ themselves are viewed as components 
of a ``magnetic vector''
\begin{equation} 
  \bi{m} = \sum_s m_s\bfeta_s \equiv \sum_i m_i \hat{\bi{h}}_i,
\label{msmi} 
\end{equation}
the vectors $\bfeta_s$ and $\hat{\bi{h}}_i$ (the latter pointing in the
direction perpendicular to the plane of rotation identified by $H_i$ in 
\sref{CoordTrafo}) being themselves interrelated by 
integer (or half-integer) coefficients.
    
An analogue of the spherical harmonics equation defining the range of $m$,
\begin{equation} 
  (l_x\pm\rmi l_y)Y_{l,\pm l}(\theta,\varphi)\equiv 
(l_x\pm\rmi l_y)^{l+1} P_l(\cos \theta)=0,
\label{SO3highest} 
\end{equation}
is formulated for a hyperspherical harmonic 
$Y_{\lambda_s\bfeta_s}(\hat{\bi{R}})$, whose degree is highest 
(i.e., can be raised no further by the ladder operator $E_{\bfeta_s})$, 
in the form 
\begin{equation} 
  E_{\bfeta_s}Y_{\lambda_s\bfeta_s}(\hat{\bi{R}}) = 0.
\label{HighestDef} 
\end{equation}
When $E_{\bfeta_s}$ is cast as a first-order differential operator,
\eref{HighestDef} determines that particular ``highest weight'' harmonic.

Operating on each of these ``highest weight'' harmonics with successions of
``lowering'' $E_{-\bfeta_s}$ operators generates complete sets of
hyperspherical harmonics, as detailed in the following sections.
\subsection{Operations on the $m_i$ parameters\label{ActOnMi}}
The $m_i$ quantum numbers, eigenvalues of the operators $H_i$, belong in
the $\hat{\bi{h}}_i$ ``space'', being raised or lowered in value by 
$E_{\balpha}$ or
$E_{\pm\bfeta_s}$ operators \emph{external} to that space, just as the $m$
eigenvalues of $l_z$ are shifted by $l_x \pm\rmi l_y$ operators with axes
orthogonal to $\hat{\bi{z}}$. As the ladder operators $l_x \pm\rmi l_y$ of 
physical 
space are viewed as ``eigenvectors'' of $l_z$ through the commutator
equations $[l_z,l_x\pm\rmi l_y] = \pm (l_x\pm\rmi l_y)$, sets of ladder
operators $E_{\balpha}$ are defined as solutions of the 
commutator-equations' set,
\begin{equation} 
  [H_i,E_{\balpha}] = \alpha_i E_{\balpha},
\qquad\alpha_i=\pm 1\textrm{\ or\ } 0;\qquad\textrm{for\ } 
i = 1, 2,\ldots\ell,
\label{LadderDef} 
\end{equation}
implying that $H_iE_{\balpha}u_i=(m_i+\alpha_i)E_{\balpha}u_i$ for
the eigenvector $u_i$ of $H_i$ with eigenvalue $m_i$.

Basic elements for solving \eref{LadderDef} emerge from the introduction and
properties of infinitesimal rotation operators  \eref{CartGen}
and \eref{SONGen} in \sref{CoordTrafo}: 
\begin{itemize}
\item[a)] Commutators $\left[H_i,J_{jk}^{xy}\right]$ 
\emph{vanish} unless one of the variables $(x_j,y_k)$ is
common to both operators. If one is, the commutator equals a related 
$J_{j'k'}^{x'y'}$ times a \emph{unit-magnitude} coefficient, much as the 
three-dimensional commutator of $l_z$ with $l_x$ or $l_y$ does.
\item[b)] Accordingly each $E_{\balpha}$ reduces to a linear combination 
  of a \emph{few} $J_{jk}^{xy}$ with $|\alpha_i| = 1$ or $0$; 
  $\alpha_i = \pm 1$ thus corresponds to raising or lowering $m_i$.
The commutator of a pair of Hermitian-conjugate ladder operators 
$[E_{\balpha},E_{-\balpha}]$ is itself Hermitian, specifically a multiple 
of the unit operator. 
\item[c)] The directions of $\balpha$ (or $\bfeta_s$) vectors in the
$\hat{\bi{h}}_i$ space run in that space at equal distances between pairs of 
$\hat{\bi{h}}_i$'s.
\item[d)] For $d$ even, the resulting operators $E_{\balpha}$ 
 (or $E_{\bfeta_s}$) 
  shift the values of \emph{two} $m_i$ parameters by unity simultaneously, 
  thus \emph{preserving the parity} of the $\sum_i m_i$. The same holds for
  most of the ladder operators for odd-dimensional systems, too, with the 
  following exception: Odd-dimensional systems include a single operator 
  $E_{\bfeta_s}$ (one among a sub-set of $\ell$ operators $E_{\balpha}$)
 acting on a single $m_i$, whose contribution violates 
  the parity conservation.
\end{itemize} 
\ref{ConstrLadder} outlines a procedure to identify raising and lowering
operators.

These elements, developed originally in \cite{Cartan,Racah}, afford a basis 
for constructing complete sets of hyperspherical harmonics by 
\emph{transforming} a \emph{single} prototype harmonic with 
\emph{complete sets} of
$E_{\balpha}$ (or $E_{\bfeta_s}$) operators.  Preferred prototypes
have a \emph{single non-zero} value of $m_i$ parameters, typically
$\{m_1=\lambda,\; m_{j\neq 1}=0\}$. An initial $E_{\balpha}$ operator
will lower $m_1$'s value by one unit, raising that of one
$|m_{j\neq 1}|$ from $0$
to $1$. Alternative successions of analogous operators generate thus
complete sets of $\lambda$-degree harmonics,
all of whose vectors $\{\bi{m}\}$ prove compatible with the initial
$\bi{m}\equiv \{\lambda,0,0,..0\}$, i.e., with $\sum_i |m_i|\leq\lambda$
and (in the even-dimensional case) $(-1)^{\sum_i m_i}=(-1)^\lambda$.
An initial prototype harmonic convenient 
for this purpose reads, in the notation of \eref{HSatt},
\begin{equation}  
  \cos^{\lambda}\alpha \sin^{\lambda}\theta_1  \rme^{\rmi\lambda
\varphi_1} \equiv \left(\frac{x_1+\rmi y_1}{R}\right)^{\lambda},  
\label{HighestFct}  
\end{equation}
with ``highest weight'' exponents $\lambda$ corresponding to 
$l_1 = m_1 = \lambda,\; l_2 = m_2 = 0, n=0$ in \eref{HSatt}.

The allowed range of each among the several $\bi{m}$ components $m_i$ is
limited by the condition $|m_i| \leq \lambda_i$, each $\lambda_i$ being an
element of the partition
\begin{equation}  
  \lambda = \sum_i \lambda_i, \qquad \lambda_i \geq 0,
\qquad -\lambda_i \leq m_i \leq \lambda_i, 
\label{lamimi} 
\end{equation}
which reflects, in turn, specific sequences of the relevant
$E_{\balpha}$ applications. Two (or more) of the $\bi{m}$ 
``weight-vectors''
thus generated may coincide, being labelled ``multiple weights'', reached by
alternative operator sequences equivalent in this respect. The total number of 
alternative $\bi{m}$ vectors generated by the present procedure, denoted by 
$w$ in \sref{HypHarm}, includes combined contributions of multiple weights. 
These $w$ 
values, representing the dimension of the accessible $\{\bi{m}\}$ set and of 
its corresponding set of hyperspherical harmonics, are discussed for 
$d$-dimensional rotations in \cite[section~10.2]{Rau} and for various other 
groups in \cite{Wybourne,Cornwell}.

The harmonics' symmetry under rotation reversals, discussed in 
\sref{LadderOps}, implies 
for even-dimensional spaces (where the parity conservation mentioned
in item d) above holds strictly) the relationship of hyperspherical harmonics
\begin{equation}  
  Y_{\lambda,-\bi{m}}(\hat{\bi{R}}) = (-1)^{\lambda}Y^*_{\lambda
\bi{m}}(\hat{\bi{R}}).    
\label{conjug}  
\end{equation}
\subsection{Operations on the $m_s$ parameters\label{ActOnMs}}
\Eref{msmi} shows parallel expansions of a ``weight-vector'' $\bi{m}$ into the 
eigenvalues $\{m_i\}$ and $\{m_s\}$ of the commuting operators $\{H_i\}$ and
of the ladder-operator labels $\{\bfeta_s\}$. These alternative base 
sets span 
the same $\ell$-dimensional space with different orientations and different 
metric scales set, respectively, by the $H_i$ eigenvalues $m_i$ with unit 
spacing and by the $\bfeta_s$ vectors with (in general) \emph{two} 
unit-magnitude 
components and resulting squared-magnitudes $\bfeta_s\cdot\bfeta_s=2$.
\Sref{ActOnMi} has stressed how ladder operators shift \emph{pairs} of $m_i$ 
parameters simultaneously, contrasting with $E_{\pm\bfeta_s}$'s shift of 
a \emph{single} $m_s$ parameter. The simpler action on the $m_s$ thus 
simplifies the construction of hyperspherical harmonics' sets.

This simplification is partly compensated by a restriction imposed on
$\{m_s\}$ sets by their equivalence to $\{m_i\}$, implied by \eref{msmi} and
represented by requiring
\begin{equation} 
  2\frac{\bi{m}\cdot\bfeta_s}{\bfeta_s\cdot
\bfeta_s} = r_s - q_s,   
\label{RootString}  
\end{equation}
whose non-negative integers $r_s$ and $q_s$ specify respectively how many
times \emph{in direct succession} the operators $E_{-\bfeta_s}$ and 
$E_{\bfeta_s}$ can be applied to $\bi{m}$. [For details on this equation, 
see section~10.1 of \cite{Wybourne}, or  sections~13.5 and 15.2 of 
\cite{Cornwell}.]    

Two critical elements underlie \eref{RootString}: (i) \Eref{msmi} establishes a
linear relationship between the parameter sets $\{m_i\}$ and $\{m_s\}$; 
(ii) The last relation in \eref{lamimi} restricts the range of each $m_i$ 
sharply. \Eref{RootString} restricts 
then the ladder operators' $E_{\pm\bfeta_s}$ action on \eref{HighestFct} or
on any hyperspherical harmonic, shifting the relevant $m_s$ value by one unit,
through a sharp \emph{selection rule}: The resulting value of $m_s$ \emph{must}
satisfy \eref{RootString}.

Within this framework one constructs complete sets of hyperspherical
harmonics in $d$ dimensions by:
\begin{itemize}
\item[a)] Selecting a coordinate system according to \sref{CoordTrafo};
\item[b)] Constructing an appropriate set of $\ell$ commuting operators 
  $\{H_i\}$;
\item[c)] Constructing corresponding sets of $\ell$ ladder operators 
$\{E_{\bfeta_s}\}$ and 
$\{E_{-\bfeta_s} \equiv E_{\bfeta_s}^{\dagger}\}$, 
in first-order differential form;
\item[d)] Determining a ``highest weight'' hyperspherical harmonic, 
  characterized by
a vector $\blambda$ with components $\lambda_i$ in the $\{H_i\}$ basis
and $\lambda_s$ in the equivalent $\{\bfeta_s\}$ basis, by solving the
relevant \eref{HighestDef}. [The prototype ``highest weight'' harmonic 
\eref{HighestFct} pertains to
$\blambda=\{\lambda,0,0,\ldots\}$ in the $\{H_i\}$ basis.];
\item[e)] Applying to this ``highest weight'' harmonic (sequences of) the 
  lowering operators $E_{-\bfeta_s}$, with $1\le s\le\ell$, for a total of 
  $2\lambda_s$ times each, to yield the succession of harmonics 
  $Y_{\lambda_sm_s}(\hat{\bi{R}})$,
terminating at $m_s = -\lambda_s$, remaining however 
\emph{within the constraints} of 
\eref{RootString} which often prevents lowering one $m_s$ value ahead of other 
$m_{s'}$'s. 
\end{itemize}
\subsection{Sample derivation of harmonics' sets
\label{ExampleSO6}}
We apply here the procedure just outlined to the three-particle system of 
\sref{JacobiPol}, forming an $SO(6)$ geometry, utilizing the same 
 notation and thus
taking care of the prescription items a) to c), except for selecting a set of
$\bfeta_s$ vectors. This set, with components indicated by $\alpha_i$ in
\eref{LadderDef}, is conventionally \cite{Cornwell} chosen as
\begin{equation} 
  \bfeta_1 \sim \{1,-1,0\},
\qquad \bfeta_2 \sim \{0,1,-1\}, 
\qquad \bfeta_3 \sim \{0,1,1\},
\label{ExBasis} 
\end{equation}
thus implicitly relating each $m_s$ number to the $m_i$ in \eref{msmi}. 
Note how the 
first of these vectors is not orthogonal to the following orthogonal pair, 
and the set of three is linearly independent.

We set the value of $\lambda$ in \eref{HighestFct} at 2 for simplicity, 
thus fixing the $\lambda_s$ values in \eref{lambdaPart} at $\{2,1,1\}$. 
The relevant Cartesian ladder operators corresponding to the set of 
$\{H_i\}$'s \eref{CartGen} take the 
form
\numparts 
\begin{eqnarray}
\eqalign{
\fl 
E_{\pm\bfeta_1} = -\rmi\left[\left(x_1\pm\rmi y_1\right)
\frac{\partial}{\partial (x_2 \pm\rmi y_2)}
-\left(x_2\mp\rmi y_2\right)
\frac{\partial}{\partial (x_1 \mp\rmi y_1)}
\right] \\ 
\lo=  \frac{1}{2}\left[ J_{12}^{xx}+J_{12}^{yy}\mp\rmi
\left(J_{12}^{xy}-J_{12}^{yx}\right)\right]
} 
\label{SO6GenCart}\\ 
\eqalign{
\fl 
E_{\pm\bfeta_2} 
= -\rmi\left[\left(z_1\mp\rmi z_2\right)
\frac{\partial}{\partial (x_2\mp\rmi y_2)}
-\left(x_2\pm\rmi y_2\right)
\frac{\partial}{\partial (z_1\pm\rmi z_2)}
\right] \\ 
\lo= \frac{1}{2}\left[J_{12}^{zx}+J_{22}^{zy}
\pm\rmi\left(J_{12}^{zy}-J_{22}^{zx}\right)\right]. 
} \\  
\eqalign{
\fl 
E_{\pm\bfeta_3} 
= -\rmi\left[\left(z_1\pm\rmi z_2\right)
\frac{\partial}{\partial (x_2\mp\rmi y_2)}
-\left(x_2\pm\rmi y_2\right)
\frac{\partial}{\partial (z_1\mp\rmi z_2)}
\right] \\ 
\lo= \frac{1}{2}\left[J_{12}^{zx}-J_{22}^{zy}
\pm\rmi\left(J_{12}^{zy}+J_{22}^{zx}\right)\right] 
} 
\end{eqnarray} 
\endnumparts 
Introducing, for the sake of compactness, the elementary operator notations,
\numparts 
\begin{eqnarray}
K^{(i)}_{\pm} = \cos\theta_i\frac{\partial}{\partial\theta_i} 
\pm\rmi\frac{1}{\sin\theta_i}\frac{\partial}{\partial \varphi_i}, 
\label{ExAux1}\\ 
L^{(i)}_{\pm} = \rme^{\pm\rmi\varphi_i}\left(\pm\frac{\partial}{\partial 
\theta_i} +\rmi\cot\theta_i\frac{\partial}{\partial\varphi_i}\right),\qquad 
i = 1,2; 
\label{ExAux2} 
\end{eqnarray} 
\endnumparts 
we recast (\ref{SO6GenCart}--$c$) in polar coordinates, corresponding to
$\{H_i\}$'s of \eref{HSGen}:
\numparts 
\begin{eqnarray} 
\fl 
E_{\pm\bfeta_1} =  -\frac{\rmi}{2} \rme^{\pm\rmi(\varphi_1-\varphi_2)} 
\left(\sin\theta_1 \sin\theta_2 \frac{\partial}{\partial\alpha} - \tan\alpha 
\sin\theta_2 K^{(1)}_{\pm} + \cot\alpha \sin\theta_1 K^{(2)}_{\mp} \right) 
\label{SO6GenHS} 
\\ 
\eqalign{
\fl 
E_{\pm\bfeta_2} = -\frac{\rmi}{2} \left[\rme^{\pm\rmi\varphi_2}\left(
\cos\theta_1 \sin\theta_2 \frac{\partial}{\partial\alpha} +\tan\alpha 
\sin\theta_1 \sin\theta_2 \frac{\partial}{\partial\theta_1} +\cot\alpha 
\cos\theta_1 K^{(2)}_{\pm}\right)\right. \\ \left.
\vphantom{\frac{\partial}{\partial\theta_1}} 
- \rmi L^{(2)}_{\pm} \right]} 
\\ 
\eqalign{
\fl 
E_{\pm\bfeta_3} =  -\frac{\rmi}{2} \left[\rme^{\pm\rmi\varphi_2}\left(
\cos\theta_1 \sin\theta_2 \frac{\partial}{\partial\alpha} +\tan\alpha 
\sin\theta_1 \sin\theta_2 \frac{\partial}{\partial\theta_1} +\cot\alpha 
\cos\theta_1 K^{(2)}_{\pm}\right)\right. \\ \left.
\vphantom{\frac{\partial}{\partial\theta_1}} 
+ \rmi L^{(2)}_{\pm} \right]} 
\end{eqnarray}
\endnumparts 

The prototype harmonic \eref{HighestFct} with $\lambda=2$ reads
as $\cos^2\alpha\sin^2\theta_1\rme^{\rmi 2\varphi_1}$, with $\lambda_i$
and $\lambda_s$ components $\{2,0,0\}$ and $\{2,1,1\}$, respectively.
Operators $E_{-\bfeta_s}$ can be applied to this expression directly,
with full attention to the condition \eref{RootString}. This condition 
excludes at the outset acting on this harmonic with 
either lowering operator $E_{-\bfeta_2}$ or $E_{-\bfeta_3}$, which 
change $m_{i=2}$ and $m_{i=3}$ from their initial 0 value without affecting 
$m_{i=1}$ that retains its highest value 2, thus violating the limitations 
on the $m_i$; only $E_{-\bfeta_1}$ operates on the harmonic 
\eref{HighestFct}
correctly. Sets of harmonics are grouped into ``layers'' according to the 
number of lowering operators that have been applied to the ``highest weight'' 
harmonic in the process. Successive steps of lowering the $m_s$ quantum 
numbers are similarly restricted, but afford alternative actions of lowering 
operators, as displayed in \tref{Tab1}. We actually show the 
first five layers of harmonics, with the $\sum_s m_s \geq 0$, the harmonics 
with negative values being obtained from these by complex conjugation 
according to \eref{conjug}.

\Tref{Tab1} also illustrates the effect of non-vanishing commutators between
ladder operators: different ordering of the same ladder operators may
yield \emph{different} harmonics with the \emph{same} label $\bi{m}$, i.e.,
a \emph{degenerate} eigenvalue, such as $\bi{m}=\{0,0,0\}$ in \tref{Tab1}.
Furthermore, 
while the procedure demonstrated in \tref{Tab1} will give the correct number
of \emph{linearly independent} harmonics, these may not necessarily be
\emph{orthogonal} (with the usual definition of a Hermitian scalar product 
as an integral over the relevant space: $(f,g)=\int\,f^*\,g\,\rmd \bi{x}$).
For a more detailed discussion of the issues of multiple eigenvalues and
orthogonalization of harmonics in the general case, see 
e.g.~\cite{Cornwell,Heim}.

Finally, recasting $\{H_i\}$ from \eref{CartGen} as 
\begin{equation}
\fl 
\eqalign{
H_i=-\rmi \left[(x_i+\rmi y_i)\frac{\partial}{\partial(x_i+\rmi y_i)}-
(x_i-\rmi y_i)\frac{\partial}{\partial(x_i-\rmi y_i)}\right],\qquad i=1,2; \\ 
H_3=-\rmi \left[(z_1+\rmi z_2)\frac{\partial}{\partial(z_1+\rmi z_2)}-
(z_1-\rmi z_2)\frac{\partial}{\partial(z_1-\rmi z_2)}\right], }
\end{equation}
and using (\ref{SO6GenCart}--$c$), we note that manipulating harmonics by means
of ladder operators is most easily performed in generic Cartesian coordinates,
where the harmonics take the form of products with generic factors 
$[(x_i\pm\rmi y_i)/R]^{|m_i|}$ \cite{Heim}. 
 
\afterpage{
\begin{landscape} 
\begin{table}
\caption{Hyperspherical harmonics for a three-particle system ($d=6$) and 
generalized angular momentum $\lambda=2$. The functions are constructed
by acting on the ``highest weight'' harmonic 
$(\cos\alpha\sin\theta_1\exp[\rmi\varphi_1])^\lambda$ with different sequences
of ladder operators, as indicated in the third column. The first two columns
list the $m$ quantum numbers in the $\bfeta_s$ and $H_i$ bases, respectively,
while the last column identifies the harmonics as \emph{superpositions} of 
functions \eref{HSatt} by listing their corresponding labels 
$\{n,l_1,m_1,l_2,m_2\}$. Indentation in the the last two columns indicates a 
continuation line of a single entry.\label{Tab1}}
\begin{tabular}{@{}lllll}
\br 
   $m_s$ & $m_i$ & Operator sequence & Resulting harmonic, not normalized
 & $\{n,l_1,m_1,l_2,m_2\}$ \\ \mr 
   $\{2,1,1\}$ & $\{2,0,0\}$ &   & 
   $\cos^2\alpha\sin^2\theta_1\rme^{\rmi2\varphi_1}$ & $\{0,2,2,0,0\}$ \\ 
\mr 
   $\{1,1,1\}$ & $\{1,1,0\}$ & $E_{-\bfeta_1}$  
 & $2\rmi\cos\alpha\sin\theta_1
    \rme^{\rmi\varphi_1}\sin\alpha\sin\theta_2\rme^{\rmi\varphi_2}$ 
 & $\{0,1,1,1,1\}$ \\ 
\mr 
   $\{0,1,1\}$ & $\{0,2,0\}$ 
 & $E_{-\bfeta_1}E_{-\bfeta_1}$ 
 & $-2\sin^2\alpha\sin^2\theta_2\rme^{\rmi 2\varphi_2}$ 
 & $\{0,0,0,2,2\}$ \\ \ms 
   $\{1,0,1\}$ & $\{1,0,1\}$ 
 & $E_{-\bfeta_2}E_{-\bfeta_1}$  
 & $2\cos^2\alpha\sin\theta_1\cos\theta_1\rme^{\rmi\varphi_1}  
    +2\rmi\cos\alpha\sin\theta_1\rme^{\rmi\varphi_1}\sin\alpha\cos\theta_2 $  
 & $\{0,2,1,0,0\}$, $\{0,1,1,1,0\}$ \\ \ms  
   $\{1,1,0\}$ & $\{1,0,-1\}$ 
 & $E_{-\bfeta_3}E_{-\bfeta_1}$  
 & $2\cos^2\alpha\sin\theta_1\cos\theta_1\rme^{\rmi\varphi_1} 
    -2\rmi\cos\alpha\sin\theta_1\rme^{\rmi\varphi_1}\sin\alpha\cos\theta_2 $ 
 & $\{0,2,1,0,0\}$, $\{0,1,1,1,0\}$ \\
\mr 
   $\{1,0,0\}$ & $\{1,-1,0\}$ 
 & $E_{-\bfeta_3}E_{-\bfeta_2}E_{-\bfeta_1}$ 
 & $2\rmi\cos\alpha\sin\theta_1\rme^{\rmi\varphi_1}\sin\alpha\sin\theta_2
   \rme^{-\rmi\varphi_2}$ 
 & $\{0,1,1,1,-1\}$  \\ \ms 
   $\{0,1,0\}$ & $\{0,1,-1\}$ 
 & $E_{-\bfeta_1}E_{-\bfeta_3}E_{-\bfeta_1}$ 
 & $2\rmi\cos\alpha\cos\theta_1\sin\alpha\sin\theta_2 \rme^{\rmi\varphi_2}   
    +2\sin^2\alpha\sin\theta_2\cos\theta_2\rme^{\rmi\varphi_2}$ 
 & $\{0,1,0,1,1\}$, $\{0,0,0,2,1\}$ \\ \ms 
   $\{0,0,1\}$ & $\{0,1,1\}$ 
 & $E_{-\bfeta_1}E_{-\bfeta_2}E_{-\bfeta_1}$ 
 & $2\rmi\cos\alpha\cos\theta_1\sin\alpha\sin\theta_2 \rme^{\rmi\varphi_2}  
    -2\sin^2\alpha\sin\theta_2\cos\theta_2 \rme^{\rmi\varphi_2}$ 
 & $\{0,1,0,1,1\}$, $\{0,0,0,2,1\}$ \\
\mr 
   $\{0,-1,1\}$ & $\{0,0,2\}$   
 & $E_{-\bfeta_2}E_{-\bfeta_1}E_{-\bfeta_2}E_{-\bfeta_1}$ 
 & $\frac{2}{3}[ \cos^2\alpha(3\cos^2\theta_1-1)
     -\sin^2\alpha(3\cos^2\theta_2-1)] $ 
 & $\{0,2,0,0,0\}$, $\{0,0,0,2,0\}$,  \\ \ms & &
 & $\quad +\frac{2}{3}\cos 2\alpha 
          +4\rmi\cos\alpha\cos\theta_1\sin\alpha\cos\theta_2$ 
 & $\quad\{1,0,0,0,0\}$,  $\{0,1,0,1,0\}$ \\ \ms 
   $\{0,0,0\}$ & $\{0,0,0\}$ 
 & $E_{-\bfeta_1}E_{-\bfeta_3}E_{-\bfeta_2}E_{-\bfeta_1}$  
 & $-\frac{2\rmi}{3}[\cos^2\alpha(3\cos^2\theta_1-1)  
    -\sin^2\alpha(3\cos^2\theta_2-1)]$ 
 & $\{0,2,0,0,0\}$, $\{0,0,0,2,0\}$, \\ \ms & & 
 & $\quad +\frac{4\rmi}{3}\cos 2\alpha$ 
 & $\quad \{1,0,0,0,0\}$ \\ \ms
   $\{0,0,0\}$ & $\{0,0,0\}$ 
 & $E_{-\bfeta_3}E_{-\bfeta_1}E_{-\bfeta_2}E_{-\bfeta_1}$ 
 & $\frac{2}{3}[\cos^2\alpha(3\cos^2\theta_1-1) 
    +2\sin^2\alpha(3\cos^2\theta_2-1)]$ 
 & $\{0,2,0,0,0\}$, $\{0,0,0,2,0\}$, \\  \ms & & 
 & $\quad +\frac{2}{3}\cos 2\alpha$ 
 & $\quad\{1,0,0,0,0\}$ \\ \ms  
   $\{0,1,-1\}$ & $\{0,0,-2\}$ 
 & $E_{-\bfeta_3}E_{-\bfeta_1}E_{-\bfeta_3}E_{-\bfeta_1}$ 
 & $\frac{2}{3}[\cos^2\alpha(3\cos^2\theta_1-1) 
    -\sin^2\alpha(3\cos^2\theta_2-1)] $ 
 & $\{0,2,0,0,0\}$, $\{0,0,0,2,0\}$, \\ \ms & & 
 & $\quad +\frac{2}{3}\cos 2\alpha  
    -4\rmi\cos\alpha\cos\theta_1\sin\alpha\cos\theta_2$ 
 & $\quad\{1,0,0,0,0\}$, $\{0,1,0,1,0\}$ \\ 
\br 
\end{tabular}
\end{table}
\end{landscape} 
}
\subsection{Harmonics for application to atoms and molecules
\label{HarmAtMol}}
Different sets of hyperspherical harmonics serve to represent different
aggregates, at different stages of their development. More specifically,
each set reflects the structure of the relevant ``Jacobi tree'' introduced
in \sref{GegenbPol}, whose ``growth'' mirrors the integration of the relevant
\eref{SEq}, as a function of its hyper-radius $R$. We deal here
with particular aspects of harmonics selection, beginning with the remark
that, at low $R$ values where centrifugal potentials prevail, governing the
single particles, the set described in \sref{JacobiPol} may prove 
adequate, complemented possibly by the set of \sref{GegenbPol}.
The following sections deal with two particular aspects of our
subject.
\subsubsection{Symmetrized harmonics\label{SymmHarm}}
Particle aggregates include generally sub-sets of identical particles:
electrons, of course, but also atomic nuclei such as the numerous protons of
hydrocarbons. As a preliminary to the eventual requirement of anti-symmetrizing
wave functions under permutation of fermion positions, or analogous operations
on bosons, it often proves convenient to select at the outset harmonics' sets
that are invariant under permutation of all identical-particle variables
\cite{Novoselsky,Barnea}. 
This preliminary operation serves particularly to reduce the dimension of
each harmonics' set.
    
Typically each electron pair may rotate about its centre of mass only with 
orbital momenta equal to an even (odd) multiple of $\hbar$ when in a singlet 
(triplet) spin state. The same holds familiarly for the nuclear states of 
para- (ortho-)molecular hydrogen. This restriction reduces the relevant 
sets' dimensions by exponential factors when applied to large sets of 
identical particles, complemented by enforcing the corresponding symmetry 
between different particle-pairs, a more laborious procedure known as the 
selection of ``fractional parentage'' \cite{Cowan}, \cite[Chapter 8]{Rau}, 
but applied more conveniently at the outset of any calculation.
\subsubsection{Expansion of whole state representations\label{ExpWholeState}}
Expanding the whole solution $\Psi(\bi{R})$ of \eref{SEq} into
hyperspherical harmonics may also serve to illustrate the resulting
representation of a multi-particle state. The set of hyperspherical 
harmonics must, however, be complemented for this purpose by a harmonic 
function of the hyper-radius $R$, a generalization of the familiar Bessel 
functions in two dimensions and of their related ``spherical Bessel'' 
functions in three dimensions. The required harmonic functions of $R$ 
belong once again to the Bessel-function family.

The spherical Bessel function equation for three dimensions, (10.1.1) of 
\cite{Tables}, differs from the standard Bessel function equation for two 
dimensions
(9.1.1), by: (i) a coefficient $2 = d-1$ inserted before its first-derivative 
term, and (ii) its eigenvalue $n(n+1) = (n + 1/2)^2 - 1/4$ replacing the 
standard
eigenvalue $\nu^2$. Similarly, the hyperspherical Bessel function for $d$
dimensions differs by (i) a coefficient $d-1\equiv 3N-4$ inserted before its 
first-derivative term, and (ii) its eigenvalue 
$\lambda(\lambda + d-2) + (d-1)(d-3)/4 = (\lambda + (d-2)/2)^2 - 1/4$ 
replacing $\nu^2$. The resulting 
Bessel function will thus be of integer or fractional order $\lambda +(d-2)/2$,
with a  corresponding pre-factor arising from the $R^{(3N-4)/2}$ coefficient of
our equation \eref{SEq}, as anticipated in \cite{Cavagnero1}. 
\section{Hyperspherical expansion of the wave equation\label{RadExp}}
The preliminary treatment of hyperspherical harmonics in 
\sref{HypHarm} suffices
to formulate an expanded version of \eref{SEq}. On the left-hand side of this
equation, the factors $R^{(3N-4)/2}$ have been separated out to allow setting
a finite initial value of $\Psi(R,\hat{\bi{R}})$ at $R$ = 0. The 
 factors
separated out take into account the centrifugal---actually, wave-mechanical---%
potential generated in polar coordinates with small values of $R$ by 
compressing particles within short ``parallel circles''. We may then
standardize wave functions that start at $R$ = 0 with unit value in a 
\emph{single hyperspherical channel} $(\lambda_0, \bmu_0)$, 
expanding as $R$ 
increases into alternative channels $(\lambda,\bmu)$, as indicated by
\begin{equation} 
\eqalign{
\Psi_{\lambda_0\bmu_0}(\bi{R}) =
\sum_{\lambda,\bmu} F_{\lambda_0\bmu_0,\lambda\bmu}(R)
Y_{\lambda\bmu}(\hat{\bi{R}}), 
\\ 
F_{\lambda_0\bmu_0,\lambda\bmu} \rightarrow R^{\lambda_0}
\delta_{\lambda_0\bmu_0,\lambda\bmu},\quad{\rm as}\;R \rightarrow 0.
}
\label{HSexpand} 
\end{equation}

Entering the expansion \eref{HSexpand} in \eref{SEq}, and projecting the result
onto the several harmonics $Y_{\lambda\bmu}(\hat{\bi{R}})$, reduces \eref{SEq}
to the system of coupled radial Schr\"{o}dinger equations
\begin{equation}  
  \frac{\rmd^2}{\rmd R^2} F_{\lambda_0\bmu_0,\lambda\bmu}(R)
+ \sum_{\lambda',\bmu'} F_{\lambda_0\bmu_0,\lambda'
\bmu'}\langle\lambda'\bmu'|k^2(R)|
\lambda\bmu\rangle = 0, 
\label{RadEq} 
\end{equation}
with the wave-number matrix
\begin{eqnarray} 
 \fl 
\langle\lambda'\bmu'|k^2(R)|\lambda
\bmu\rangle =  
\left(2{\cal M}E-\frac{\lambda(\lambda+d-2)+[(d-2)^2-1]/4
}{R^2} \right) \delta_{\lambda'\bmu',\lambda\bmu} 
\nonumber \\ 
+\frac{2{\cal M}\langle \lambda'\bmu'|Z(\hat{\bi{R}})|\lambda\bmu\rangle}{R} 
+\ldots 
\label{wavenum} 
\end{eqnarray}
Here $Z(\hat{\bi{R}})\,/R$ represents the Coulomb potential energy of the
interacting particles, evaluated at each hyper-radius $R$. The dots at the
end of \eref{wavenum} stand for any additional terms of the $k^2$ matrix
corresponding to Hamiltonian terms that
represent spin-orbit or relativistic corrections not included explicitly. The
ability of \eref{RadEq} to include such effects---thus far not exploited---%
may by-pass the current need to treat such terms perturbatively rather than 
directly in the basic equation.

The system of coupled equations \eref{RadEq} is formally infinite, owing to the
infinite range of its parameter $\lambda$, thus seemingly impractical as noted
in \sref{Intro}. However, circumstances also described in \sref{Intro}
 reduce its size generally to a modest level.
\subsection{Displaying the evolution toward fragmentation\label{Evolve}}
Our approach to displaying this evolution derives from features of the
fragmentation of nuclei that are held together by short-range interactions 
\cite{Calogero}. 
Beyond this range, $r_0$, energy eigenfunctions resolve into 
\emph{fragmentation
eigenchannels}, labelled here by $\rho$, each of them propagating
at $r > r_0$ in force-free space with \emph{unchanged structure}, i.e., with 
spherical wave-fronts, $r$-independent angular distributions 
$f_{\rho}(\theta,\varphi)$ and uniform phases $\phi_{\rho}(r)$. The parameter 
sets $\{f_{\rho}(\theta,\varphi)\}$ and $\{\phi_{\rho}(r)\}$ embody here the 
effect of all particle interactions at $r < r_0$.

The opportunity occurred in \cite{Calogero} to utilize an analogous 
parametrization
\emph{regardless of interaction ranges}, by embodying the effect of \emph{all} 
interactions at the ranges $0\leq r \leq R$ into parameter sets 
$\{f_{\rho}(R;\hat{\bi{R}})\}$ and $\{\phi_{\rho}(R)\}$ to be evaluated 
\emph{for successive values of R}. These sets' dependence on $R$ displays 
each eigenchannel's 
evolution as $R$ increases from its 0 value (at the system's centre of mass)
towards $\infty$, i.e., disregarding all interactions at $r > R$ at each step 
of integration.

The hyperspherical channel functions \eref{HSexpand}, 
$\Psi_{\lambda_0\bmu_0}(\bi{R})$, of each multi-particle system serve 
here as a basis to calculate the $\{f_\rho(R;\hat{\bi{R}})\}$ 
and $\{\phi_\rho(R)\}$ parameters by casting them 
into superpositions
\begin{equation} 
\fl 
\Phi_{\rho}(\bi{R}) = \sum_{\lambda_0\bmu_0}
\langle\rho(R)|\lambda_0\bmu_0\rangle \Psi_{\lambda_0\bmu_0}(\bi{R})
= \sum_{\lambda_0\bmu_0} \langle\rho(R)|\lambda_0\bmu_0\rangle
\sum_{\lambda\bmu} F_{\lambda_0\bmu_0,\lambda\bmu}(R)
Y_{\lambda\bmu}(\hat{\bi{R}}) 
\label{ChannelFct} 
\end{equation}  
with initial values of the coefficients
\begin{equation} 
\fl 
\langle\rho(R)|\lambda_0\bmu_0\rangle \rightarrow 1, 
\qquad\textrm{i.e.,\ }\langle\rho|\rightarrow\langle\lambda_0\bmu_0|
\qquad\textrm{as\ } R \rightarrow 0.  
\label{ChannelInit}  
\end{equation}

Requiring each phase $\phi_{\rho}(R)$ of $\Phi_{\rho}(\bi{R})$, and
its gradient $(\rmd\phi/\rmd R)$, to be uniform over each hyper-surface 
($R=\mathrm{const}.$) identifies each superposition \eref{ChannelFct}, 
$\Phi_\rho(\bi{R})$, as an \emph{eigenchannel} of the propagating 
Schr\"odinger equation \eref{SEq} at 
\emph{each value} of $R$. It thus implies that the angular eigenfunction 
$f_{\rho}(R;\hat{\bi{R}})$, representing the aggregate's ``shape'' at
$R$, evolves ``in step'' from each hyper-surface to the next. This requirement
translates into a system of linear homogeneous equations for the coefficients
$\langle\rho(R)|\lambda_0\bmu_0\rangle$.

Following the ``phase--amplitude'' approach of \cite{Calogero}, which 
replaces second order wave equations with pairs of first order equations, 
reference \cite{Fano,Bohn1} introduced the eigenphase's 
$\tan\phi_{\rho(R)}$---without 
previous reference to \eref{ChannelFct}---as an eigenvalue of the ``R-matrix''
\begin{eqnarray} 
  \langle\lambda\bmu|{\cal R}(R)|\lambda'\bmu'\rangle 
 & = & 
\sum_{\lambda_0\bmu_0}\left(\frac{\rmd F}{\rmd R}\right)^{-1}_{\lambda
\bmu,\lambda_0\bmu_0} F_{\lambda_0\bmu_0,\lambda'\bmu'}(R) 
 \nonumber \\ & = & 
\sum_{\rho}\langle\lambda\bmu|\rho(R)\rangle\tan 
\phi_{\rho(R)} \langle\rho(R)|\lambda'\bmu'\rangle. 
\label{Rmatrix} 
\end{eqnarray}
Reference \cite{Bohn2} proceeded then by taking the derivative of 
\eref{Rmatrix} with respect to $R$, replacing its element 
$(\rmd^2F/\rmd R^2)$ with its expression \eref{RadEq} in terms of 
the $k^2$ matrix \eref{wavenum}, and finally transforming the result
to the $\langle\rho(R)|$ basis by means of the coefficients 
$\langle\rho(R)|\lambda\bmu\rangle$ and their reciprocals. 

Taking the ${\cal R}$ matrix as a stepping-stone served thus to resolve 
the initial \emph{second-order} equation \eref{RadEq} into the set of 
\emph{first-order equations} corresponding to the diagonal and off-diagonal 
elements of \eref{Rmatrix}, respectively,
\numparts
\begin{eqnarray} 
\fl 
\frac{\rmd\tan\phi_{\rho(R)}}{\rmd R} = 1 + 
\langle\rho(R)|k^2(R)|\rho(R)\rangle\;\tan^2\phi_{\rho(R)}, 
\label{PhaseEqA} \\ 
\fl 
\frac{\rmd\langle\rho(R)|\lambda\bmu\rangle}{\rmd R} = 
\sum_{\rho' \neq \rho} \frac{\sin\phi_{\rho(R)}\langle\rho(R)|k^2(R)|
\rho'(R)\rangle \sin\phi_{\rho'(R)}}{\sin(\phi_{\rho} -
\phi_{\rho'})} \langle\rho'(R)|\lambda\bmu\rangle, 
\label{PhaseEqB}  
\end{eqnarray}
\endnumparts
with $R$ expressed in units of $(\rmd R/\rmd\phi_{\rho})$. 
This set of equations
has been integrated numerically in \cite{Bohn2} for the prototype example
of doubly excited He, with conclusions described below.
\subsection{Illustration and discussion\label{Discussion}}
The numerical integration of (\ref{PhaseEqA}--$b$) for He has 
been carried out
with a device that accelerates the convergence of expansion into harmonics
$\langle \lambda\bmu|$, by replacing these harmonics with eigenvectors
of the $k^2$ matrix \eref{wavenum} at each $R$. These eigenvectors represent 
``adiabatic'' solutions of our problems, carried out earlier in the frame 
of \cite{Macek1,Macek2}, i.e., disregarding the coupling between the radial and 
angular variables, $R$ and $\hat{\bi{R}}$.

The sample results shown in figure~1 should be viewed as interconnecting 
each system's sets of compact $(c)$ and fragmented $(f)$ channels, outlined 
in \cite{Fano81} and \cite{Bohn2}. Quantum mechanical scattering theory
represents this connection by ``Jost'' matrices $J_{fc}$ \cite{Newton}. Each 
standing-wave eigenfunction of \eref{SEq}, identified by an initial boundary 
condition $(c)$ at its compact limit, is represented asymptotically near its 
$(f)$ limit by $\sum_f \sin(k_f R) J_{fc}$, or more conveniently in terms of 
its outgoing and incoming components
\begin{equation} 
  \sum_f \exp(\rmi k_f R) J^+_{fc},\qquad\sum_f\exp(-\rmi k_f R)J^-_{fc}.  
  \label{Jost} 
\end{equation}

The matrices $J^{\pm}_{fc}$ are complex-conjugate for ``open'' channels 
$f$, i.e., when energy suffices to achieve the $f$ limit; for energetically 
``closed'' channels the wave-number $k_f$ is imaginary, whereby one component 
converges to zero at discrete eigenvalues of the energy $E$ and the other 
diverges as $R \rightarrow \infty$. The Jost matrices serve then to construct 
scattering matrices
\begin{equation}  
  S_{f'f} = \sum_c J^+_{f'c} \left(\frac{1}{J^-}\right)_{cf}, 
  \label{Smatrix}
\end{equation}
as detailed in \cite{Fano81,Fano83a}. In this frame one views each amplitude
$\langle \rho(R)| \lambda_0\bmu_0\rangle$ of \eref{ChannelFct}, 
evaluated at a finite range $R$, as a partial construction of the Jost 
matrix element $J_{fc}$
with $f$ representing the limit of $\langle\rho(R)|$ as $R\rightarrow \infty$.
Plots of the several bra symbols $\langle \rho(R)|$ achieve our objective
of displaying the system's evolution from $R= 0$ toward $\infty$.

\begin{figure}
\begin{center}
\includegraphics[width=62mm]{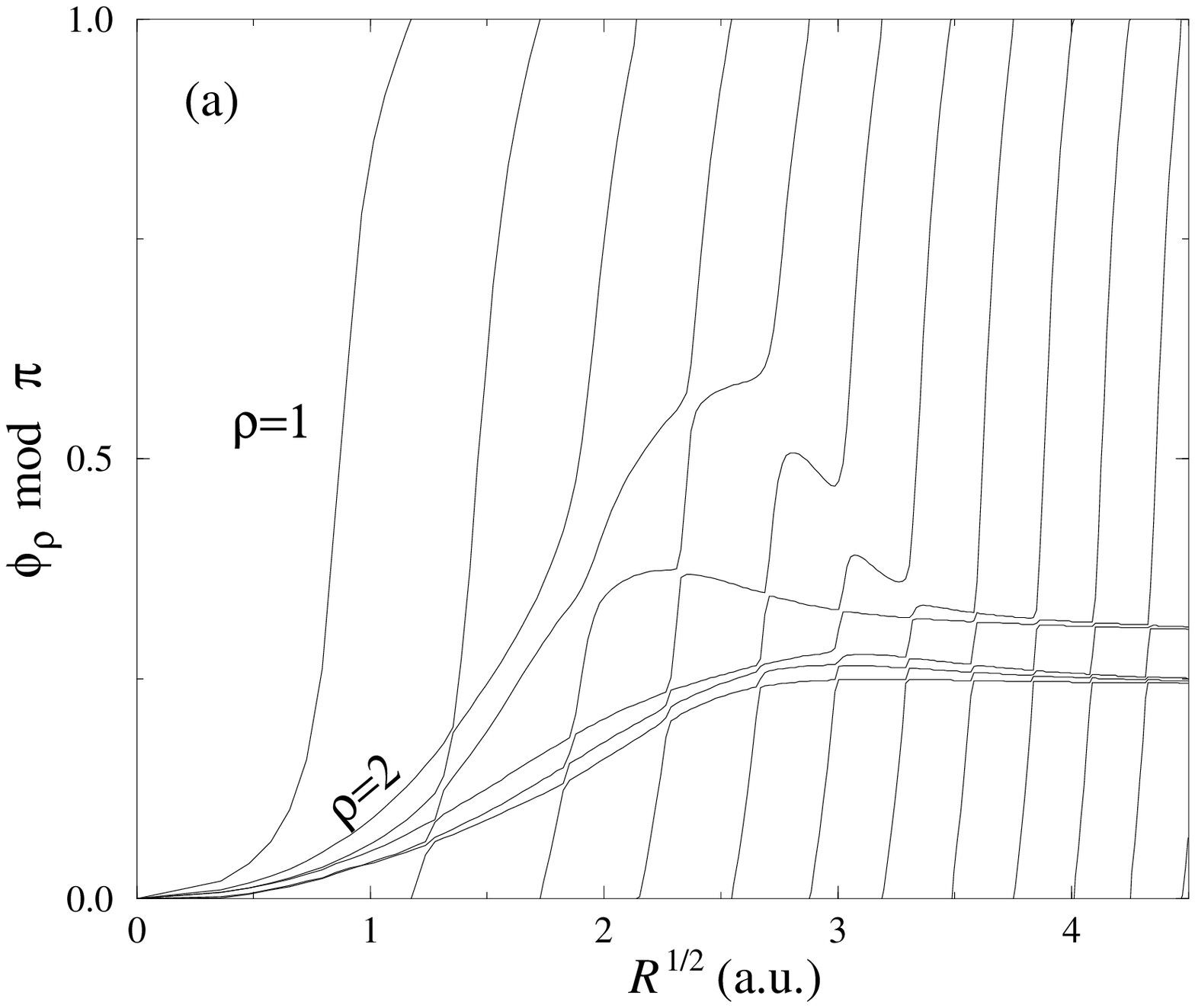}\hfill
\includegraphics[width=62mm]{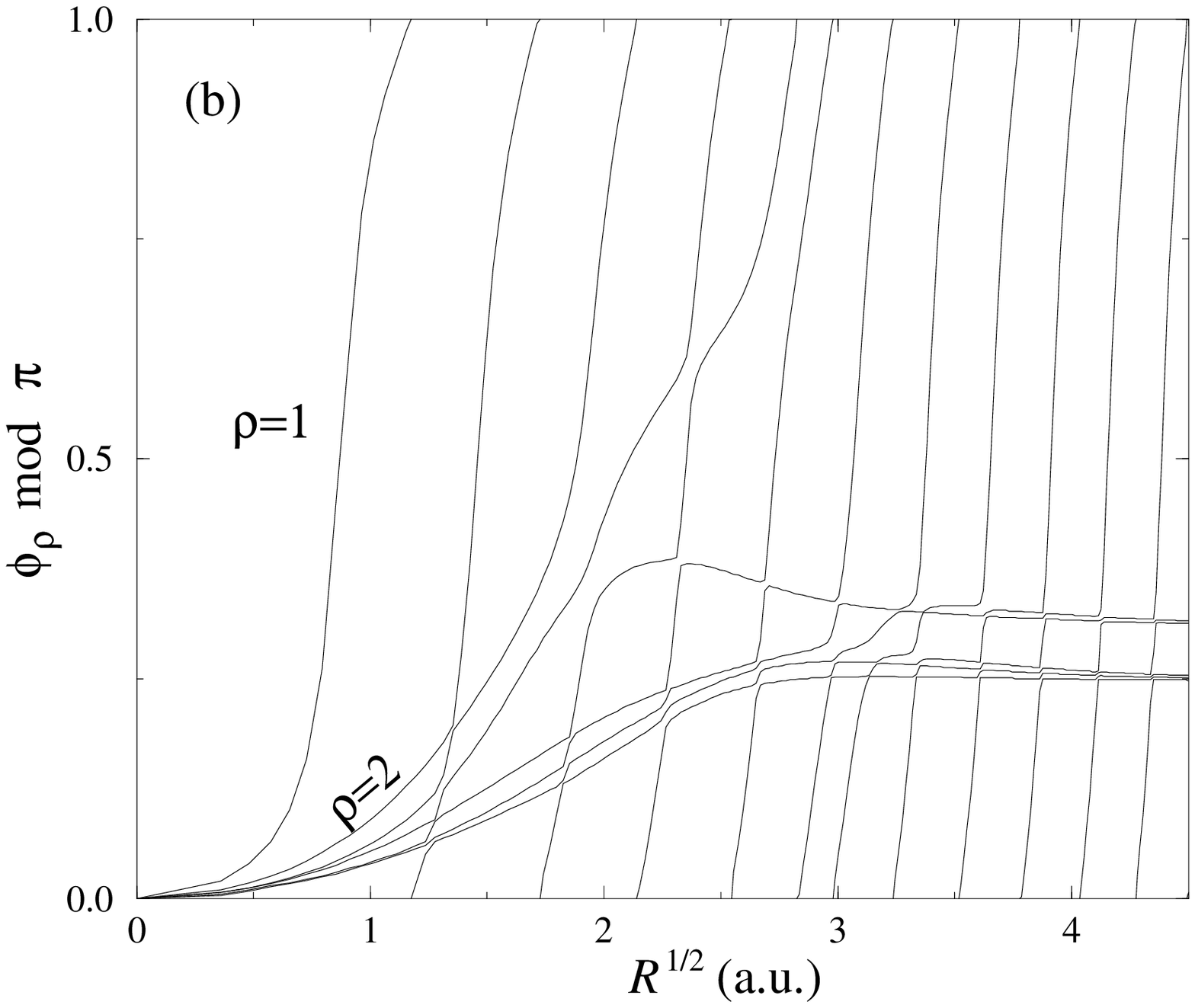}
\end{center}
\caption{(a) Eigenphases $\phi_\rho$ modulo $\pi$, vs the square root of the
hyper-radius, calculated at the total energy $E=57$ eV above the ground state,
just below the ``$(2s^2)\, ^1S$'' resonant state in helium. (b) shows the
same set of eigenphases, calculated at a total energy $E=58$ eV above the
ground state.\label{Fig1}}
\end{figure}

Figures~\ref{Fig1}(a) and (b) plot eigenphases $\phi_{\rho}(R)$ vs.\ 
$\sqrt{R}$, modulo
$\pi$, at energies straddling the $(2s)^2\;^1S$ resonance of He near 58~eV, for
a number of $\langle\rho(R)|\lambda_0\bmu_0\rangle$ pairs. Each line's 
slope mirrors the rate of increase of $\phi_{\rho}(R)$, i.e.\ (loosely), the 
rate of expansion in $R$ of the corresponding eigenfunction. [The first 
eigenphase's slope reflects the rapid motion of an electron ionized with 
approximately 33~eV kinetic energy; successive curves reflect the increasingly
slower development of two-electron excitations in successively higher modes.] 
In the lower and upper ranges of the ordinate $\phi_{\rho}$, corresponding to 
low values of $\sin[\phi_{\rho}(R)]$, pairs of curves appear to cross with 
minimal disturbance, owing to unresolved values of the coupling coefficient on 
the right of \eref{PhaseEqB}, in spite of the singularity arising from the 
vanishing 
of its denominator at each crossing. Major effects of crossings emerge instead
at middle ranges of $\phi_{\rho}(R)$, where pairs of curves appear to repel 
each other experiencing major deflexions.

The localization of such major ``avoided crossings'' within limited ranges
of ordinates and abscissas, confirmed by analogous more dense plots, verifies
our expectation that eigenfunctions propagate smoothly outside the limited
parameter ranges where the coupling coefficients of \eref{PhaseEqB} diverge. 
The 
apparent ``repulsion'' of curves at avoided crossings is a familiar aspect of 
the Landau-Zener phenomenon \cite{Landau,Zener,Stuck} occurring at points of 
``degeneracy'' where two alternative values of a parameter
coincide, as the phases $(\phi_{\rho},\;\phi_{\rho'})$ do here (modulo $\pi$).

An additional major manifestation of avoided crossings, namely, the
hybridization of pairs
$(\langle\rho|\lambda_0\bmu_0\rangle$, 
$\langle \rho'| \lambda_0'\bmu_0'\rangle)$ 
has, however, not been included in the calculations underlying 
figure~\ref{Fig1}. This hybridization might result by fitting 
parameters of the observed phenomena---slopes, slope differences of the
curves and their closest approaches---to the corresponding elements of the 
Landau-Zener theory. The relevance of the present 
$\langle \rho|\lambda_0\bmu_0\rangle$ coefficients to the Jost matrices 
of interest remains 
fragmentary pending further developments of Nakamura's recent
analytical improvements on the Landau-Zener theory \cite{Zhu1,Zhu2}.
\subsection{Qualitative effects of Coulomb interactions\label{Coulomb}}
\Eref{wavenum}'s matrix 
$\langle\lambda'\bmu'|Z(\hat{\bi{R}})|\lambda\bmu\rangle$ consists of
$N(N-1)/2$ terms for an aggregate of
$N$ charged particles. These terms, included sequentially in computer
programs, serve to solve \eref{PhaseEqA} and \eref{PhaseEqB} numerically, 
yet warrant analysis 
aimed at visualizing their qualitative action. To this end one may resolve the
$Z$-matrix' action on the wave function $\Psi$ into its several aspects.

Sub-sets of terms acting between particles with equal (opposite) charges
push corresponding particle sub-sets apart (together). Within this scope one
notes that:
\begin{itemize}
\item[a)] Expressing the distance between each atomic nucleus and another
particle in terms of their common mass-weighted hyper-radius $R$ 
\emph{scales up} its charge by its mass' square root, thereby boosting its 
interactions' strength and thus favouring \emph{molecular dissociation} as 
compared to ionization by electron ejection.
\item[b)] The force acting on each particle-pair depends on the pair's
orientation. Combining the Coulomb forces between various particle-pairs
involves thus extensive geometrical transformations of the relevant position
coordinates.
\end{itemize}
Coulomb interactions between sub-sets of \emph{identical particles} may be
conveniently grouped, particularly so following the symmetrization of relevant
position coordinates outlined in \sref{HarmAtMol} and in 
 \ref{Antisymmetry}. It would then become possible to combine 
Coulomb terms pertaining to such sub-sets to yield expressions of their 
electrical multipole moments and of their corresponding multipole fields. A
semi-macroscopic view of each aggregate's mechanics should thus emerge, in
terms of collective variables.

Within the context of initial combination of molecular nuclei with closed-shell
electrons, one could then envisage treating all molecular valence electrons as 
forming a gas whose electric multipole moments are inflated by electronic
mutual repulsions, yet contained by the attractive multipole fields of nuclei
and closed-shell cores. The proton-nuclei of hydrocarbon molecules would be
similarly viewed. This attractive containment would perform a two-fold action:
Holding the molecule together as a unit and simultaneously smoothing out the
distribution of opposite charges throughout its volume.
\appendix
\renewcommand{\numparts}{\addtocounter{equation}{1}%
     \setcounter{eqnval}{\value{equation}}%
     \setcounter{equation}{0}%
     \def\theequation{\ifnumbysec
     \Alph{section}.\arabic{eqnval}{\it\alph{equation}}%
     \else\arabic{eqnval}{\it\alph{equation}}\fi}}
\renewcommand{\endnumparts}{\def\theequation{\ifnumbysec
     \Alph{section}.\arabic{equation}\else
     \arabic{equation}\fi}%
     \setcounter{equation}{\value{eqnval}}}
\section{Construction of ladder operators\label{ConstrLadder}}
The generic infinitesimal operators \eref{SONGen} raise and lower the 
eigenvalues,
$m_i$ and $m_j$, of operator pairs $(H_i,H_j)$ by mapping a single harmonic
onto a \emph{superposition} of harmonics with different $m_i$'s and $m_j$'s.
We construct here linear
combinations of those operators that raise \emph{or} lower by unity these
eigenvalues in a definite way, according to \eref{LadderDef}.

Casting for this purpose $(H_i,H_j)$ in the form of the first two expressions
in \eref{CartGen}, i.e., as $(H_i\equiv J_{ii}^{xy}, H_j\equiv J_{jj}^{xy})$, 
restricts the solutions $E_{\balpha}$ of \eref{LadderDef} to combinations
of the four components
\begin{equation}
J_{ij}^{xx},\qquad J_{ij}^{yy},\qquad J_{ij}^{xy},\qquad J_{ij}^{yx}.
\label{SO6Rot}
\end{equation}
Identification of the proper linear combinations 
proceeds through analysis of their commutator relations with the $H_i$'s,
namely,
\numparts 
\begin{eqnarray}
[H_i,J_{ij}^{xx}]=+\rmi J_{ij}^{yx}, \qquad  
[H_j,J_{ij}^{xx}]=+\rmi J_{ij}^{xy},  \\ {}
[H_i,J_{ij}^{yx}]=-\rmi J_{ij}^{xx}, \qquad   
[H_j,J_{ij}^{xy}]=-\rmi J_{ij}^{xx},  \\ {}
[H_i,J_{ij}^{yy}]=-\rmi J_{ij}^{xy}, \qquad  
[H_j,J_{ij}^{yy}]=-\rmi J_{ij}^{yx},  \\ {}
[H_i,J_{ij}^{xy}]=+\rmi J_{ij}^{yy}, \qquad  
[H_j,J_{ij}^{yx}]=+\rmi J_{ij}^{yy}. 
\end{eqnarray}
\endnumparts 
Combinations of the operators \eref{SO6Rot} symmetric and antisymmetric in 
their $(x,y)$ variables, analogues of \eref{ijpm}'s symmetry in $(i,j)$ 
indices, yield
\numparts 
\begin{eqnarray}
E_{\pm\balpha_{ij}^{(++)}} = (J_{ij}^{xx}-J_{ij}^{yy})
\pm \rmi(J_{ij}^{xy}+J_{ij}^{yx}),  \label{SO6Ladpp} \\
E_{\pm\balpha_{ij}^{(+-)}} = (J_{ij}^{xx}+J_{ij}^{yy})
\mp \rmi(J_{ij}^{xy}-J_{ij}^{yx}), \label{SO6Laddpm} 
\end{eqnarray}
\endnumparts 
satisfying the desired equations
\numparts 
\begin{eqnarray}
[H_i,E_{\pm\balpha_{ij}^{(++)}}]=\pm E_{\pm\balpha_{ij}^{(++)}}, 
\qquad 
[H_j,E_{\pm\balpha_{ij}^{(++)}}]=\pm E_{\pm\balpha_{ij}^{(++)}}, 
 \\ {}
[H_i,E_{\pm\balpha_{ij}^{(+-)}}]=\pm E_{\pm\balpha_{ij}^{(+-)}}, 
\qquad 
[H_j,E_{\pm\balpha_{ij}^{(+-)}}]=\mp E_{\pm\balpha_{ij}^{(+-)}},  
\end{eqnarray}
\endnumparts 
with
\numparts 
\begin{eqnarray}
\balpha_{i,j}^{(++)}=(+1,+1), \\ 
\balpha_{i,j}^{(+-)}=(+1,-1), \\ 
\balpha_k^{(++)}=\balpha_k^{(+-)}=0,\qquad\textrm{for\ } k\ne i,j. 
\end{eqnarray}
\endnumparts 
Note that the two operator sets 
$\{\frac{1}{2}(J_{ij}^{xx}\mp J_{ij}^{yy}),\;
\frac{1}{2}(J_{ij}^{xy}\pm J_{ij}^{yx}),\; \frac{1}{2}(H_j\pm H_i)\}$
commute exactly like $\{l_x,l_y,l_z\}$ in three dimensions. The occurrence of
(symmetric or antisymmetric) \emph{pairs} of rotation operators in the
role of both $l_x$ and $l_y$, respectively, reflects the feature of the
pairwise change of $m_i$ and $m_j$. The symmetry of these operator pairs 
under interchange of the \emph{first and second coordinates} 
$x_i\leftrightarrow y_i$ and $x_j\leftrightarrow y_j$ of the relevant 
$H_i$ and $H_j$ extends the present construction to ladder operators that 
change the $m$ quantum numbers of $H$ operators of the third type in 
\eref{CartGen}, involving two $z$-coordinates.

For odd dimensions, $\ell$ ladder operator pairs pertaining to coordinate
pairs $(i,0)$ change only one
of the $m_i$'s. If again $x_i, y_i$ make up the operator $H_i$ pertaining to
$m_i$, the two ladder operators acting on $m_i$ are complex linear 
combinations
of the two infinitesimal rotation operators which involve either of $x_i$ or
$y_i$, together with the single unpaired coordinate [denoted $x_0$ in 
\sref{CoordTrafo}, 
e.g. the $z$ coordinate in the familiar case of $SO(3)$].
In this case, the ladder operators are completely analogous to the ones in
$SO(3)$.
\section{Transformation between Jacobi trees\label{TrafoJacobi}}
As an example of transformation between Jacobi trees, consider the two
trees

\begin{center}
\setlength{\unitlength}{10pt}
\begin{picture}(16,10)
\put(1,7){\line(1,-2){2}}
\put(2,7){\line(-1,-2){0.5}}
\put(3,7){\line(1,-2){0.5}}
\put(4,7){\line(-1,-2){1.5}}
\put(5,7){\line(-1,-2){2.0}}
\put(3,1.5){\line(0,1){1.5}}
\put(0.8,8){1}
\put(1.8,8){2}
\put(2.8,8){3}
\put(3.8,8){4}
\put(4.8,8){5}
\put(2.7,0.2){A}
\put(6,5){\vector(1,0){4.0}}
\put(11,7){\line(1,-2){2}}
\put(12,7){\line(1,-2){1.5}}
\put(13,7){\line(-1,-2){0.5}}
\put(14,7){\line(-1,-2){1.0}}
\put(15,7){\line(-1,-2){2.0}}
\put(13,1.5){\line(0,1){1.5}}
\put(10.8,8){1}
\put(11.8,8){2}
\put(12.8,8){3}
\put(13.8,8){4}
\put(14.8,8){5}
\put(12.7,0.2){D}
\end{picture}
\end{center}

The mass-weighted relative coordinates for tree A result from 
independent-particle coordinates by the transformation 
\numparts 
\begin{eqnarray}
\fl 
\bxi_1^A = \sqrt{\frac{M_1M_2}{M_1+M_2}}(\bi{r}_2 - \bi{r}_1) 
 \\
\fl 
\bxi_2^A = \sqrt{\frac{M_3M_4}{M_3+M_4}}(\bi{r}_4 - \bi{r}_3) 
\\
\fl 
\bxi_3^A = \sqrt{\frac{(M_1+M_2)(M_3+M_4)}{M_1+M_2+M_3+M_4}}
\left(\frac{M_3\bi{r}_3+M_4\bi{r}_4}{M_3+M_4}
-\frac{M_1\bi{r}_1+M_2\bi{r}_2}{M_1+M_2} \right)   \\
\fl 
\bxi_4^A = \sqrt{\frac{(M_1+M_2+M_3+M_4)M_5}{M_1+M_2+M_3+M_4+M_5}}
\left(\bi{r}_5-\frac{M_1\bi{r}_1+M_2\bi{r}_2+
M_3\bi{r}_3+M_4\bi{r}_4}{M_1+M_2+M_3+M_4}\right), 
\end{eqnarray}
\endnumparts 
i.e.\ by first connecting particles 1 and 2, then 3 and 4, then the complex
$\{12\}$ to the complex $\{34\}$, and finally the complex $\{1234\}$ to 5. 
The fifth Jacobi coordinate represents the centre of mass which  
remains the same for all trees consisting of the same particles, and is  
hence irrelevant to transformations of the four relative coordinates.

The transformation from A to D resolves into three elementary 
``transplantations'':
\begin{center}
\setlength{\unitlength}{10pt}
\begin{picture}(36,10)
\put(1,7){\line(1,-2){2}}
\put(2,7){\line(-1,-2){0.5}}
\put(3,7){\line(1,-2){0.5}}
\put(4,7){\line(-1,-2){1.5}}
\put(5,7){\line(-1,-2){2.0}}
\put(3,1.5){\line(0,1){1.5}}
\put(0.8,8){1}
\put(1.8,8){2}
\put(2.8,8){3}
\put(3.8,8){4}
\put(4.8,8){5}
\put(2.7,0.2){A}
\put(6,5){\vector(1,0){4.0}}
\put(7,6){$T_{AB}$}
\put(11,7){\line(1,-2){2}}
\put(12,7){\line(1,-2){1.0}}
\put(13,7){\line(1,-2){0.5}}
\put(14,7){\line(-1,-2){1.5}}
\put(15,7){\line(-1,-2){2.0}}
\put(13,1.5){\line(0,1){1.5}}
\put(10.8,8){1}
\put(11.8,8){2}
\put(12.8,8){3}
\put(13.8,8){4}
\put(14.8,8){5}
\put(12.7,0.2){B}
\put(16,5){\vector(1,0){4.0}}
\put(17,6){$T_{BC}$}
\put(21,7){\line(1,-2){2}}
\put(22,7){\line(1,-2){1.0}}
\put(23,7){\line(-1,-2){0.5}}
\put(24,7){\line(-1,-2){1.5}}
\put(25,7){\line(-1,-2){2.0}}
\put(23,1.5){\line(0,1){1.5}}
\put(20.8,8){1}
\put(21.8,8){2}
\put(22.8,8){3}
\put(23.8,8){4}
\put(24.8,8){5}
\put(22.7,0.2){C}
\put(26,5){\vector(1,0){4.0}}
\put(27,6){$T_{CD}$}
\put(31,7){\line(1,-2){2}}
\put(32,7){\line(1,-2){1.5}}
\put(33,7){\line(-1,-2){0.5}}
\put(34,7){\line(-1,-2){1.0}}
\put(35,7){\line(-1,-2){2.0}}
\put(33,1.5){\line(0,1){1.5}}
\put(30.8,8){1}
\put(31.8,8){2}
\put(32.8,8){3}
\put(33.8,8){4}
\put(34.8,8){5}
\put(32.7,0.2){D}
\end{picture}
\end{center}
The transformation $T_{AB}$ from A to B affects only the first and the
third among the mass-weighted Jacobi coordinates, since it 
transplants branch 2 from the complex $\{12\}$ to the complex $\{234\}$.
It is therefore represented by applying to the four-vector 
$(\bxi_1^A,\bxi_2^A,\bxi_3^A,\bxi_4^A)$ the matrix
\numparts 
\begin{equation}
T_{AB} = \left(\begin{array}{cccc}
\cos\phi_{AB} & 0 & -\sin\phi_{AB} & 0 \\
0 & 1 & 0 & 0 \\
\sin\phi_{AB} & 0 & \cos\phi_{AB} & 0 \\
0 & 0 & 0 & 1 \end{array}\right), 
\end{equation}
i.e.\ by a ``kinematic rotation'' through an angle 
\begin{equation}
\phi_{AB} = \tan^{-1}\sqrt{\frac{M_2(M_1+M_2+M_3+M_4)}{M_1(M_3+M_4)}}
\end{equation}
\endnumparts 
according to \eref{rotJacAngle}. Note the general structure in the mass
coefficients: Transplanting branch $q$ from the complex $\{pq\}$ to the
complex $\{qr\}$ corresponds to a rotation by 
$\phi=\tan^{-1}\sqrt{M_q(M_p+M_q+M_r)/M_pM_r}$ (in the first quadrant, i.e.\
with positive signs for both cosine and sine). 
The transformation $T_{BC}$ transplants 
branch 3 from $\{34\}$ to $\{23\}$, and $T_{CD}$ transplants
the complex branch $\{234\}$ from $\{1234\}$ to $\{2345\}$, with the respective
transformation matrices 
\numparts 
\begin{equation}
\eqalign{
T_{BC} = \left(\begin{array}{cccc}
\cos\phi_{BC} & -\sin\phi_{BC} & 0 & 0 \\
\sin\phi_{BC} &  \cos\phi_{BC} & 0 & 0 \\
0 & 0 & 1 & 0 \\ 0 & 0 & 0 & 1 \end{array}\right) 
\\ T_{CD} = \left(\begin{array}{cccc}
1 & 0 & 0 & 0 \\ 0 & 1 & 0 & 0 \\ 
0 & 0 & \cos\phi_{CD} & -\sin\phi_{CD} \\
0 & 0 & \sin\phi_{CD} &  \cos\phi_{CD} 
\end{array}\right) 
}
\end{equation}
and rotation angles
\begin{equation}
\fl 
\eqalign{
\phi_{BC} = \tan^{-1}\sqrt{\frac{M_3(M_2+M_3+M_4)}{M_2M_4}}  \\
\phi_{CD} = \tan^{-1}\sqrt{\frac{(M_2+M_3+M_4)(M_1+M_2+M_3+M_4+M_5)}{
M_1M_5}}.
}
\end{equation}
\endnumparts 
The complete transformation for this sequence 
is represented by $T_{AD} = T_{CD}\cdot T_{BC}\cdot T_{AB}$ with  
\begin{equation}
\fl 
T_{AD}
= \left(\begin{array}{cccc} 
\cos\phi_{AB}\cos\phi_{BC} & -\sin\phi_{BC} & -\sin\phi_{AB}\cos\phi_{BC} & 0
\\
\cos\phi_{AB}\sin\phi_{BC} & \cos\phi_{BC} & -\sin\phi_{AB}\sin\phi_{BC} & 0 \\
\sin\phi_{AB}\cos\phi_{CD} & 0 & \cos\phi_{AB}\cos\phi_{CD} & -\sin\phi_{CD} \\
\sin\phi_{AB}\sin\phi_{CD} & 0 & \cos\phi_{AB}\sin\phi_{CD} & \cos\phi_{CD} 
\end{array}\right) 
. 
\end{equation}
Inserting the explicit expressions for the angles $\phi$ verifies
that the coordinates $\bxi_i^D,\, 1\le i\le 4,$ 
indeed describe the relative coordinates of tree D in terms of 
independent particles coordinates, namely
\numparts 
\begin{eqnarray}
\fl 
\bxi_1^D = \sqrt{\frac{(M_2+M_3)M_4}{M_2+M_3+M_4}}\left(
\frac{M_2\bi{r}_2+M_3\bi{r}_3}{M_2+M_3}-\bi{r}_4\right)  \\
\fl 
\bxi_2^D = \sqrt{\frac{M_2M_3}{M_2+M_3}}(\bi{r}_2-\bi{r}_3) 
 \\
\fl 
\bxi_3^D = \sqrt{\frac{(M_2+M_3+M_4)M_5}{M_2+M_3+M_4+M_5}}\left(
\frac{M_2\bi{r}_2+M_3\bi{r}_3+M_4\bi{r}_4}{M_2+M_3+M_4}-\bi{r}_5\right) 
 \\
\fl 
\bxi_4^D = \sqrt{\frac{M_1(M_2+M_3+M_4+M_5)}{M_1+M_2+M_3+M_4+M_5}}
\left(\frac{M_2\bi{r}_2+M_3\bi{r}_3+M_4\bi{r}_4+M_5\bi{r}_5}{M_2+M_3+M_4
+M_5}-\bi{r}_1\right). 
\end{eqnarray} 
\endnumparts 
\section{Finite Transformations of Hyperspherical Harmonics.
\label{TrafoHarmonics}}
Finite transformations of multi-dimensional harmonics (or operators) 
correspond to the infinitesimal ones considered in \sref{Classify} just as the
prototype transformation \eref{lzef} (pertaining to a physical-space rotation)
corresponds to the infinitesimal \eref{lz}. This correspondence holds 
generally,
since all transformations relevant to this paper resolve into products of
two-dimensional rotations, as stressed repeatedly in the text.

\Sref{Classify} has identified hyperspherical harmonics in the frame of a
representation based on a vector $\blambda$ in the $\ell$-dimensional
space of maximally commuting operator sets $\{H_i\}$. Within this scope, we
might deal here just with $\ell$-dimensional rotations of $\blambda$. This
space itself is, however, subject to rotations of the $\{H_i\}$ induced by the
$3(N-1)$-dimensional coordinate rotations considered in \sref{CoordTrafo} for
$N$-particle aggregates. Generic infinitesimal operators on  such spaces were
indicated in \eref{SONGen} by $J^{xy}_{ij}$, whose label $xy$ refers to a 
pair of coordinate axes, whereas $ij$ refers to a pair of particles.

Viewing physical-space rotations, identified by three Euler angles, as our
model, recall how two of these angles pertain to rotations about a 
$z$-axis (hence parallel to an $xy$-plane) and the third one to a shift of 
$\hat{\bi{z}}$'s orientation to a new direction $\hat{\bi{z}}'$, 
 usually understood to 
lie on the previous $xz$-plane. Whereas rotations by an angle $\varphi$ about 
$\hat{\bi{z}}$ simply multiply eigenvectors of $l_z$ with eigenvalue $m$ by 
$\rme^{\rmi m\varphi}$, rotations by an angle $\theta$ in the $xz$-plane 
transform it 
into a superposition of eigenvectors whose eigenvalues $m'$ result by
transforming the initial $m$ with the Wigner matrix $d^{(l)}_{m'm}(\theta)$.
[The index $l$ stands here for the largest value (``highest weight'') of $m$.]

Correspondingly, in a multi-dimensional space, we consider two distinct
classes of two-dimensional rotations: 
(i) Rotations about one of the $\{H_i\}$ operators'
symmetry axes (i.e., in a plane orthogonal to that axis) which multiply a
harmonic eigenvector of $H_i$ with eigenvalue $|m_i| \leq \lambda_i$ by
$\rme^{\rmi m_i\varphi_i}$; and (ii) Orientation changes of an $H_i$'s own 
axis, within
a specified plane through that axis, yielding a superposition of harmonics
with Wigner coefficients $D^{\lambda_i}_{\bmu',\bmu}$, whose
subscripts differ only by replacing their $m_i$ component with $m_i'$.

The multi-dimensional framework deals with transforming from one coordinate
\emph{basis}, including its operator set $\{H_i\}$, to a new \emph{basis}
with its operator set $\{H_j'\}$, each set of indices $\{i=1,2,\ldots\}$
and $\{j=1,2,\ldots\}$ being ordered. This framework affords articulating 
generic transformations through sequences of two-dimensional rotations by 
Euler angles $\{\varphi_{\alpha}\}$ and $\{\theta_{\nu}\}$:
The initial $\varphi_{\alpha=1}$ equals the angle between the $H_i$'s
zero-azimuth and the plane of the pair $\{H_{i=1},H'_{j=1}\}$'s axes. The 
initial $\theta_{\nu=1}$ equals similarly the angle between the axes of
$H_{i=1}$ and $H'_{j=1}$. The next $\varphi_{\alpha=2}$ shifts the plane of
the $H_{i=1},H'_{j=1}$ axes to $H'_{j=1}$'s zero-azimuth. Corresponding angles
$\{\varphi_{\alpha=3}, \theta_{\nu=2}, \varphi_{\alpha=4}\}$ pertain to the
operator pair $\{H_{i=2},H'_{j=2}\}$, a procedure to continue recursively.

Insofar as the $\{H_i\}$ operators are anchored to their coordinate
systems, their two classes of rotations drag their coordinate axes along.
Altogether, transformations of harmonics indices by coordinate rotations
are thus seen to resolve into three elements: (i) Rotation of the
``representation vector'' $\blambda$ in the $\ell$-dimensional space of
the $\{H_i\}$ set; (ii) Rotation of the $\{H_i\}$'s themselves, described 
above; (iii) Further rotations of coordinates with respect to the $\{H_i\}$. 
\subsection*{Example: Kinematic rotation of harmonics}
We transform here hyperspherical harmonics of two Jacobi vectors representing
a three-particle system. Equations (\ref{rotJac}) and \eref{rotJacAngle} 
have described the transformation of a Jacobi vector pair from tree $A$ 
to tree $B$
\numparts 
\begin{eqnarray}
A:\qquad \left\{\begin{array}{l}
\bxi_1^A=\sqrt{\frac{M_1M_2}{M_1+M_2}}(\bi{r}_1-\bi{r}_2) \\
\bxi_2^A=\sqrt{\frac{(M_1+M_2)M_3}{M_1+M_2+M_3}}(\frac{M_1\bi{r}_1
+M_2\bi{r}_2}{M_1+M_2}-\bi{r}_3)\end{array}\right. \label{Tree1} \\
B:\qquad \left\{\begin{array}{l}
\bxi_1^B=\sqrt{\frac{M_2M_3}{M_2+M_3}}(\bi{r}_2-\bi{r}_3) \\
\bxi_2^B=\sqrt{\frac{M_1(M_2+M_3)}{M_1+M_2+M_3}}(\bi{r}_1-
\frac{M_2\bi{r}_2+M_3\bi{r}_3}{M_2+M_3}),\end{array}\right. \label{Tree2}
\end{eqnarray}
\endnumparts 
as a kinematic rotation 
by an angle $\beta=\tan^{-1}\sqrt{(M_1+M_2+M_3)M_2/(M_1M_3)}$,
\begin{equation}
\{\bxi_1^B,\; \bxi_2^B\} =\{\cos\beta\bxi_1^A-\sin\beta
\bxi_2^A,\; \sin\beta\bxi_1^A+\cos\beta\bxi_2^A\}.
\label{vecrot12}
\end{equation}
Since each vector $\bxi$ has three spatial components, we deal here with
a six-dimensional coordinate transformation with components
\begin{equation}
\fl 
\{\bxi_1^A,\;\bxi_2^A\}=\{(\bxi_1^A)_x,(\bxi_1^A)_y,(\bxi_1^A)_z,
  (\bxi_2^A)_x,(\bxi_2^A)_y,(\bxi_2^A)_z\}\equiv\{x_1,y_1,z_1,x_2,y_2,z_2\},
\label{coordrot12}
\end{equation}
and similarly for $\bxi_i^B$. 
The $x$-component of $\bxi_1^B$
results from a rotation by the angle $\beta$ in the $x_1x_2$-plane, 
the $x$-component of $\bxi_2^B$ by a rotation through $-\beta$ 
in the (oriented) $x_2x_1$-plane, and likewise for their 
$y$- and $z$-components. 
The corresponding transformation of three-particle harmonics 
$|\lambda,\bmu_A\rangle$, as represented, e.g., by \eref{HSatt}, 
is indicated according to \sref{HypHarm} by 
\begin{equation}
|\lambda\bmu_B\rangle = T_{AB}(\beta)|\lambda\bmu_A\rangle 
= \sum_{\bmu_A'}|\lambda\bmu_A'\rangle D_{\bmu_A',
\bmu_A}^\lambda(\beta).
\label{RotHarm}
\end{equation}
Dealing here with Cartesian coordinate rotations of the $\bxi$ vectors,
at variance with the preceding polar coordinate description, our present
operator $T_{AB}$ factors into three separate (commuting) transformations of
harmonics corresponding to rotations by $\beta$ in each of the three coordinate
planes $x_1x_2$, $y_1y_2$, and $z_1z_2$ 
\begin{equation}
T_{AB}(\beta)=\exp(\rmi\beta J_{12}^{xx})\exp(\rmi\beta J_{12}^{yy})
\exp(\rmi\beta J_{12}^{zz}).
\label{TrafoGen}
\end{equation}
Convenient expressions of the three infinitesimal operators in this expression
of $T_{AB}$ appear in earlier parts of this paper: (i) \Eref{CartGen} 
identifies
$J_{12}^{zz}$ as the operator $H_3$ for our three-particle system, yielding 
immediately
\begin{equation}
\exp(\rmi\beta J_{12}^{zz})|\lambda\bmu_A\rangle = \exp(\rmi m_3\beta)
|\lambda\bmu_A\rangle; 
\label{efJzz}
\end{equation}
(ii) \Eref{SO6Laddpm}
identifies the combination $J_{12}^{xx}+J_{12}^{yy}$ as sum of the operator
$E_{\balpha_{12}^{(+-)}}$ and its reciprocal 
$E_{-\balpha_{12}^{(+-)}}$, 
both acting on the eigenvalues $m_1$ and $m_2$ of the $\{H_1,H_2\}$
pair, the symbol $\balpha_{12}^{(+-)}$ meaning ``raising $m_1$ and 
lowering $m_2$''. The combination 
$E_{\balpha_{12}^{(+-)}}+E_{-\balpha_{12}^{(+-)}}$
of two reciprocal (Hermitian conjugate) operators is itself Hermitian. 
The resulting matrix elements 
$\langle\lambda\bmu_A'|\exp(\rmi\beta\{E_{\balpha_{12}^{(+-)}}
+E_{-\balpha_{12}^{(+-)}}\})|\lambda\bmu_A\rangle$ 
amount to Wigner $d_{m',m}^{(j)}$ elements, with
parameters specified by the following observations.
Projecting $\bmu_A$ and $\bmu_A'$ onto 
$\balpha_{12}^{(+-)}$ generalizes the lower indices in $d_{m',m}^{(j)}$
to the present higher-dimensional setting:
\begin{equation}
m = \frac{\bmu_A\cdot\balpha_{12}^{(+-)}}{
\balpha_{12}^{(+-)}\cdot\balpha_{12}^{(+-)}} = \frac{1}{2}(m_1-m_2),
\label{generalM}
\end{equation}
and similarly $m'=(m_1'-m_2')/2$. \Sref{ActOnMs} discussed in item~e) 
 a succession, or \emph{chain}, of harmonics labelled here by 
$\bmu_A+n\balpha_{12}^{(+-)}$, 
with total length $2\lambda_s$, restricted however by \eref{RootString}.
The ``multipole order'' represented by the upper parameter in $d_{m',m}^{(j)}$ 
corresponds here to half the length of this chain of harmonics. Finally, the 
generalization of the ``triangular condition'' familiar from three dimensions
now requires $|\lambda\bmu_A\rangle$ and $|\lambda\bmu_A'\rangle$ to
lie on the \emph{same} chain of harmonics 
$|\lambda,\bmu_A+n\balpha_{12}^{(+-)}\rangle$. 
Analogy to the relation $l_x=\frac{1}{2}(l_++l_-)$ suggests treating 
$\exp[\rmi\beta(E_{\balpha_{12}^{(+-)}}+E_{-\balpha_{12}^{(+-)}})]$
like a rotation about the $x$-axis in three dimensions. With the standard
definition of Euler-angle rotations, the corresponding matrix element picks 
up an additional phase factor $\exp[\rmi (m'-m)\pi/2]$ resulting from 
rotating the $y$- onto the $x$-axis and back. Note also that the sum of 
reciprocal operators in our matrix element lacks the factor $1/2$,
thereby effectively multiplying the angle $\beta$ in the $d$-symbol's
argument by a factor 2. The procedure, outlined here for a three-particle 
example, extends similarly to larger aggregates.
\section{Outline of procedures for treating large sets of identical particles
\label{Antisymmetry}}
Constructing wave functions of a few (3--4) electrons, with the required
anti-symmetrization, is rather familiar, being extended to atomic 
shell-filling, e.g., in Chapter~8 of \cite{Rau}. Its extension
to much larger sets remains problematic. Deceptively simple considerations,
to be presented below, indicate that this extension may actually proceed along 
the same lines, essentially because successive steps prove independent of one 
another. Whether these considerations constitute more than just a ``solution
in principle'' for this fundamental problem of quantum many-body theory 
remains to be seen. The inevitable exponential proliferation of operations 
to be carried out hampers their actual implementation for all but the 
smallest sets of particles. Significant simplifications occur particularly 
for a system's ground state configuration owing to (in general) higher 
symmetry in this state. Even for this case, however, the anti-symmetrization
of sets with more than three identical particles still presents a formidable
task. The most promising approach to this problem's systematic solution 
\emph{and its implementation} is currently being developed by Barnea and
Novoselsky \cite{Barnea} employing the concepts discussed in this
Topical Review, namely, Jacobi coordinates and hyperspherical harmonics.
\subsection*{Coordinates and their symmetrization}
Identity of particles implies that permutation of \emph{any} pair of them 
leaves any function of the pair unchanged. Artificial labelling of such 
particles, by indices $i=1,2,3,\ldots,$ appears nevertheless generally 
desirable for purposes of ``book-keeping''. 
It is then necessary to ``symmetrize'' any function of particle 
positions, $f(\bi{r}_1,\bi{r}_2,\ldots)$ to ensure its \emph{invariance 
under permutation of each pair} of indices $(i,j)$, by superposing 
sets of such functions differing by the whole set of relevant permutations. 
Spin and position coordinates should be combined, of course, in this 
construction that proves increasingly laborious with increasing number of 
particles.  
\subsection*{Classification by seniority} 
The ``seniority'' label (``$v$'') of an atomic state indicates the number of 
its particle pairs characterized as $^1S$ and thus isolated from its remaining
particles. This characterization means invariance under rotation of space
coordinates of both spin and position variables. The spin part of this label
applies equally regardless of the total number of particles in the system. The
``$S$'' label implies a spherically symmetric matching of the pair's angular
distribution, whose extension to multi-dimensional systems needs elaboration.
Considering that this symmetry is attained by combining a pair of orbitals
even and odd, respectively, under reflection through a plane%
---thus stretching in orthogonal directions---%
we suggest achieving the corresponding
invariance in higher dimensions by combining pairs of hyperspherical harmonics
pertaining to Jacobi trees constructed by selecting orthogonal space
directions at each step-wise addition of one particle.
\subsection*{Introduction of ``triple tensors''}
This operation, introduced by Judd in the 1960's and described in 
\cite[pp~209ff]{Rau}, yields a systematic classification of the shell-filling
process for the electrons of each atomic shell. It rests on elementary
applications of coordinate-rotation transformation pairs and of their
``reduction'', which appear as equally serviceable regardless of their
dimensionality.  
\subsection*{Separation of particle sub-sets}
This operation, familiar in atomic systems  and leading there to the
``fractional parentage'' procedure, has a major role in multi-particle settings
where, typically, electron sub-sets perform varied functions forming 
``closed shells'' (or sub-shells) of different atoms as well as chemical bonds,
\emph{preserving the relevant symmetry and coherences}. The structure and 
flexibility of the Jacobi trees corresponding to alternative hyperspherical 
harmonics appear well suited to extension to multi-particle settings, with
appropriate development of recursion techniques.

\end{document}